\documentstyle[12pt,fleqn,epsfig]{article}

\begin{document}

\begin{center}
  {\Huge \bf  Transverse Momentum Fluctuations } 
\end{center}
\begin{center}
  {\Huge \bf in Nuclear Collisions at 158 $A$GeV  } 
\end{center}

\vspace{0.5cm}

\begin{center}
  {\Large \bf The NA49 Collaboration}
\end{center}

\vspace{0.5cm}

\begin{abstract}
\noindent
Results are presented on event-by-event fluctuations in transverse 
momentum of charged particles, produced at forward rapidities in p+p, C+C, 
Si+Si and Pb+Pb collisions at 158 $A$GeV. Three different characteristics 
are discussed: the average transverse momentum of the event, the 
$\Phi_{p_{T}}$ fluctuation measure and two-particle transverse momentum 
correlations. In the kinematic region explored, the dynamical fluctuations  
are found to be small. However, a significant system size dependence of 
$\Phi_{p_{T}}$ is observed, with the largest value measured in peripheral 
Pb+Pb interactions. The data are compared with predictions of several
models.

\end{abstract}

\newpage

\begin{center}
  {\bf The NA49 Collaboration}
\end{center}
\vspace{0.5cm}
\noindent
T.~Anticic$^{21}$, B.~Baatar$^{9}$,D.~Barna$^{4}$,
J.~Bartke$^{7}$,  M.~Behler$^{14}$,
L.~Betev$^{10}$, H.~Bia{\l}\-kowska$^{19}$, A.~Billmeier$^{10}$,
C.~Blume$^{8}$,  B.~Boimska$^{19}$, M.~Botje$^{1}$,
J.~Bracinik$^{3}$, R.~Bramm$^{10}$, R.~Brun$^{11}$,
P.~Bun\v{c}i\'{c}$^{10,11}$, V.~Cerny$^{3}$, 
P.~Christakoglou$^{2}$, O.~Chvala$^{16}$,
J.G.~Cramer$^{17}$, P.~Csat\'{o}$^{4}$, N.~Darmenov$^{18}$,
A.~Dimitrov$^{18}$, P.~Dinkelaker$^{10}$,
V.~Eckardt$^{15}$, P.~Filip$^{15}$,
D.~Flierl$^{10}$,Z.~Fodor$^{4}$, P.~Foka$^{8}$, 
P.~Freund$^{15}$,
V.~Friese$^{8}$, J.~G\'{a}l$^{4}$,
M.~Ga\'zdzicki$^{10}$, G.~Georgopoulos$^{2}$, E.~G{\l}adysz$^{7}$, 
S.~Hegyi$^{4}$, C.~H\"{o}hne$^{14}$, 
K.~Kadija$^{21}$, A.~Karev$^{15}$,
V.I.~Kolesnikov$^{9}$, T.~Kollegger$^{10}$, 
R.~Korus$^{13}$, M.~Kowalski$^{7}$, 
I.~Kraus$^{8}$, M.~Kreps$^{3}$, M.~van~Leeuwen$^{1}$, 
P.~L\'{e}vai$^{4}$, L.~Litov$^{18}$, M.~Makariev$^{18}$, 
A.I.~Malakhov$^{9}$, 
C.~Markert$^{8}$, M.~Mateev$^{18}$, B.W.~Mayes$^{12}$, 
G.L.~Melkumov$^{9}$,
C.~Meurer$^{10}$,
A.~Mischke$^{8}$, M.~Mitrovski$^{10}$, 
J.~Moln\'{a}r$^{4}$, St.~Mr\'owczy\'nski$^{13}$,
G.~P\'{a}lla$^{4}$, A.D.~Panagiotou$^{2}$, D.~Panayotov$^{18}$,
K.~Perl$^{20}$, A.~Petridis$^{2}$, M.~Pikna$^{3}$, L.~Pinsky$^{12}$,
F.~P\"{u}hlhofer$^{14}$,
J.G.~Reid$^{17}$, R.~Renfordt$^{10}$, W.~Retyk$^{20}$,
C.~Roland$^{6}$, G.~Roland$^{6}$, 
M. Rybczy\'nski$^{13}$, A.~Rybicki$^{7,11}$,
A.~Sandoval$^{8}$, H.~Sann$^{8}$, N.~Schmitz$^{15}$, P.~Seyboth$^{15}$,
F.~Sikl\'{e}r$^{4}$, B.~Sitar$^{3}$, E.~Skrzypczak$^{20}$,
G.~Stefanek$^{13}$,
 R.~Stock$^{10}$, H.~Str\"{o}bele$^{10}$, T.~Susa$^{21}$,
I.~Szentp\'{e}tery$^{4}$, J.~Sziklai$^{4}$,
T.A.~Trainor$^{17}$, D.~Varga$^{4}$, M.~Vassiliou$^{2}$,
G.I.~Veres$^{4}$, G.~Vesztergombi$^{4}$,
D.~Vrani\'{c}$^{8}$, A.~Wetzler$^{10}$,
Z.~W{\l}odarczyk$^{13}$
I.K.~Yoo$^{5}$, J.~Zaranek$^{10}$, J.~Zim\'{a}nyi$^{4}$

\vspace{0.5cm}
\noindent
$^{1}$NIKHEF, Amsterdam, Netherlands. \\
$^{2}$Department of Physics, University of Athens, Athens, Greece.\\
$^{3}$Comenius University, Bratislava, Slovakia.\\
$^{4}$KFKI Research Institute for Particle and Nuclear Physics, Budapest, Hungary.\\
$^{5}$Department of Physics, Pusan National University, Busan, Republic of Korea. \\
$^{6}$MIT, Cambridge, USA.\\
$^{7}$Institute of Nuclear Physics, Cracow, Poland.\\
$^{8}$Gesellschaft f\"{u}r Schwerionenforschung (GSI), Darmstadt, Germany.\\
$^{9}$Joint Institute for Nuclear Research, Dubna, Russia.\\
$^{10}$Fachbereich Physik der Universit\"{a}t, Frankfurt, Germany.\\
$^{11}$CERN, Geneva, Switzerland.\\
$^{12}$University of Houston, Houston, TX, USA.\\
$^{13}$Institute of Physics \'Swi{\,e}tokrzyska Academy, Kielce, Poland.\\
$^{14}$Fachbereich Physik der Universit\"{a}t, Marburg, Germany.\\
$^{15}$Max-Planck-Institut f\"{u}r Physik, Munich, Germany.\\
$^{16}$Institute of Particle and Nuclear Physics, Charles University, Prague, Czech 
Republic.\\
$^{17}$Nuclear Physics Laboratory, University of Washington, Seattle, WA, USA.\\
$^{18}$Atomic Physics Department, Sofia University St. Kliment Ohridski, Sofia, 
Bulgaria.\\ 
$^{19}$Institute for Nuclear Studies, Warsaw, Poland.\\
$^{20}$Institute for Experimental Physics, University of Warsaw, Warsaw, Poland. \\
$^{21}$Rudjer Boskovic Institute, Zagreb, Croatia.\\

\newpage

\section{Introduction}

Nucleus-nucleus (A+A) collisions at relativistic energies have been
intensely studied over the last two decades. The main goal of these
efforts is to understand the properties of strongly interacting matter 
under extreme conditions of high energy and baryon densities when 
the creation of the quark-gluon plasma (QGP) is expected 
\cite{Col75}. Experimental results obtained in a broad range of 
collision energies indicate that an extended zone of strongly interacting 
dense matter indeed occurs in the collision process. Various collision 
characteristics and their collision energy dependence  suggest 
\cite{Gaz99a,na49_energy} that a transient state of deconfined
matter may be created at collision energies as low as 40 $A$GeV. 

\indent
QGP formation is expected to occur at the early collision stage when
the system is sufficiently hot and dense. In the course of further
evolution, the system dilutes and cools down, hadronizes and
finally decays into free hadrons. Therefore the final state hadrons 
carry only indirect information about the early stage of the collision.
Thus firm conclusions about the creation of deconfined matter
require studying a variety of collision characteristics.

As fluctuations are sensitive to the dynamics of the system,
in particular at the phase transition, 
the analysis of event-by-event fluctuations has been proposed as 
an important tool in the study of A+A collisions \cite{Hei00}. 
Large acceptance detectors, which allow the observation of a significant 
fraction of the final state particles, have made this suggestion truly 
attractive \cite{stock}. First pioneering studies were carried out by the 
NA49 collaboration on the fluctuations of the average transverse momentum 
\cite{App99} and the $K/ \pi $ ratio \cite{KPi} in central Pb+Pb 
collisions at the CERN SPS.

\indent
There are a number of collision characteristics which can
be inferred from experimental data on event-by-event fluctuations. 
First of all, the fluctuation analysis 
can help to resolve the long standing problem whether, or rather to 
what extent, the strongly interacting matter, emerging from the 
early collision stage, achieves both thermal \cite{Gaz92,Mro98a,Mro99a} 
and chemical \cite{Gaz99b,Mro99b} equilibrium. In the NA49 investigation 
of event-wise fluctuations in the $K/ \pi $ ratio \cite{KPi} 
in central Pb+Pb collisions at 158 $A$GeV, no significant deviations from 
the hadro-chemical equilibrium ratio were found. 
If the equilibrium is indeed reached, the system's heat capacity 
\cite{Sto95, SRS} and its compressibility \cite{Mro98b} can, in 
principle, be deduced from the temperature and multiplicity fluctuations 
respectively. While large multiplicity fluctuations can be treated as a 
signal of particle production via clusters \cite{Shur02} or super-cooled  
droplets of deconfined matter \cite{Bay99}, small final state 
fluctuations of the conserved charges, electric or baryonic, can occur 
when fluctuations generated in the plasma phase are frozen 
due to the rapid expansion of the system \cite{Jeo00,Asa00}. 
On the other hand, significant transverse momentum and
multiplicity fluctuations can result if the system hadronizes
from a QGP near the predicted second order critical QCD end point 
\cite{SRS, Anton}. The latter has been shown by recent lattice QCD 
studies to occur at a substantial baryochemical potential \cite{Fodor}, 
characteristic of the CERN SPS energy range. 

\indent
This paper extends the previous study \cite{App99} of the NA49 experiment 
of event-by-event transverse momentum fluctuations.  
The main objective is to observe how the fluctuation pattern changes with 
increasing number of nucleons participating in a collision, i.e. with the 
system size. For this purpose, not only Pb+Pb collisions at various 
centralities are studied, but also p+p, C+C and Si+Si interactions at a 
beam energy of 158 GeV per nucleon. In particular, it will be checked, 
whether the correlations present in the final state of p+p interactions 
survive in the collisions of heavier systems, as expected if the 
nucleus+nucleus (A+A) collision is a simple superposition of 
nucleon+nucleon (N+N) interactions. Moreover a test will 
be performed of the reasonable expectation that the fluctuations 
become more similar to those of an equilibrated system when 
the number of participating nucleons increases.

\indent
Measuring event-by-event fluctuations in A+A collisions
one should consider the influence of two trivial sources of
fluctuations. The first one is caused by event-by-event fluctuations 
of the collision geometry and the second one by the finite number of
particles (statistical fluctuations). The dynamical fluctuations of 
interest have to be extracted from the noise caused by these trivial sources.

\indent
In such a situation, a suitable choice of statistical tools for the study 
of event-by-event fluctuations is really important. In this work, 
mainly the fluctuation measure $\Phi$ introduced in \cite{Gaz92} will be 
employed. However, other fluctuation measures have also been proposed and 
studied, e.g. $\sigma_{p_{T},dyn}$ \cite{sigma_Volosh}, 
$\Delta \sigma_{p_{T}}$ \cite{Tra00} and $\Sigma_{p_{T}}$ 
\cite{star_2002}, which can all be related. 
$\Phi$ equals exactly zero when inter-particle correlations are 
absent. It also eliminates `geometrical' fluctuations due to the impact 
parameter variation. Thus, $\Phi$ is `deaf' to the statistical 
noise and `blind' to the collision centrality. The $\Phi$ measure
was already used in the previous experimental study \cite{App99}  
of $p_T$ fluctuations. It was also calculated within various models of 
nuclear collisions  
\cite{Bleicher:1998wu,Capella:1999uc,Liu:1998xf,Utyuzh:2001up,Korus:2001fv,Kor01}. 
However, in these model considerations, 
the effects of experimental acceptance, which significantly influence 
the results, were usually not taken into account. Consequently, most of 
these results cannot be compared to the data.

\indent
For better understanding of the structure of the correlations 
contributing to $\Phi$, two-particle transverse momentum correlations (as 
proposed in \cite{Tra00}) are studied as well. A preliminary analysis was 
presented in \cite{Reid}.

\indent
This paper is organized as follows:
In Sec. \ref{s:measures} the statistical tools used in this analysis 
are introduced and briefly discussed. The experimental set-up and 
procedures are presented in Sec. \ref{s:experiment}. 
Experimental effects such as detector acceptance and two-track
resolution are discussed in Sec. \ref{s:data}.
The results on the system dependence 
of the $p_T$ fluctuations are presented in Sec. \ref{s:results}. 
Discussion of the results and their comparison with theoretical 
models is given in Sec. \ref{s:discussion}. A summary closes the paper.

\section{Measures of fluctuations}
\label{s:measures}

There are numerous observables which can be used to measure $p_T$ 
fluctuations in high energy collisions. A natural one is the distribution 
of the average transverse momentum of the events defined as

\begin{equation}
M(p_{T})=\frac{1}{N}\sum_{i=1}^{N}p_{Ti},
\end{equation}
where $N$ is the multiplicity of accepted particles in a given 
event and $p_{Ti}$ is the transverse momentum of the $i$-th particle. 
The distribution of $M(p_{T})$ is usually compared to the 
corresponding distribution obtained for `mixed events' in which
the particles are independent from each other and follow the
experimental inclusive spectra (the multiplicity distribution for mixed 
events is the same as for the data). A difference 
between the two distributions signals the presence of dynamical 
fluctuations. Since the $M(p_{T})$ distribution crucially depends
on the particle multiplicity, the method cannot be used
to compare systems of significantly different multiplicities.

\indent
A more appropriate measure is the quantity $\Phi$  \cite{Gaz92} 
which, by its construction, is insensitive to the system size. 
In this paper the $\Phi$ measure is used for the analysis of $p_{T}$
fluctuations ($\Phi_{p_{T}}$). Following the authors of \cite{Gaz92}, one
defines the single-particle variable $z_{p_{T}}=p_{T}-\overline{p_{T}}$  
with the bar denoting averaging over the single-particle inclusive 
distribution. One easily observes that $\overline{z_{p_{T}}} = 0$.
Further, one introduces the event variable $Z_{p_{T}}$, which is a
multi-particle analog of $z_{p_{T}}$, defined as
\begin{equation}
Z_{p_{T}}=\sum_{i=1}^{N}(p_{Ti}-\overline{p_{T}}),
\end{equation}
where the summation runs over particles in a given event. Note, that 
$\langle Z_{p_{T}} \rangle = 0$, where $\langle ... \rangle$ represents 
averaging over events. Finally, the $\Phi_{p_{T}}$ measure is defined 
as

\begin{equation}
\label{Phi}
\Phi_{p_{T}}=\sqrt{\frac{\langle 
Z_{p_{T}}^{2} \rangle }{\langle N
\rangle }}-\sqrt{\overline{z_{p_{T}}^{2}}}.
\label{eq_phi}
\end{equation}

The second part of Eq.~(\ref{eq_phi}) is simply the dispersion of the 
inclusive $p_T$ distribution (further denoted as $\sigma_{p_T}$). 
It can be easily shown that $\Phi_{p_{T}} = 0$, when no inter-particle 
correlations are present and the single-particle spectrum is independent
of multiplicity. As already mentioned, $\Phi_{p_{T}}$ is insensitive to 
centrality. This property may be expressed as follows:
$\Phi_{p_{T}}$ is independent of the distribution of the number of 
particle sources if the sources are identical and independent from each other
\cite{Gaz92,Mro99a}. In particular, $\Phi_{p_{T}}$ does not depend on 
the impact parameter if the A+A collision is a simple superposition of 
N+N interactions.

\indent
In spite of the above mentioned advantages, there is an important 
disadvantage of using $\Phi_{p_{T}}$ in the fluctuation analysis. While
$\Phi_{p_{T}}$ is  sensitive to the presence of dynamical correlations
in a system, it does not disentangle their nature.
Fluctuations of very different character 
contribute to $\Phi_{p_{T}}$. In order to achieve a better 
understanding of the fluctuation structure one needs to apply a more 
differential method \cite{Tra00}. 

The correlations can be studied by plotting the cumulative $p_T$ variables
of particle pairs. Namely, for a given particle, instead of its $p_{T}$ 
one introduces the variable $x$ defined as \cite{Bia90}

\begin{equation}
x(p_T)=\int_{0}^{p_{T}}\rho({p_{T}}')d{p_{T}}',
\label{def_3}
\end{equation}
where $\rho(p_{T})$ is the inclusive $p_{T}$ distribution, normalized to
unity, which is obtained from all particles used in the 
analysis. By construction, the $x$ variable varies 
between 0 and 1 with a {\em flat} probability distribution. 
The two-particle correlation plots, as presented in this paper, are
obtained by plotting $(x_{1},x_{2})$ points for all possible particle
pairs within the same event. The number of pairs in each $(x_{1},x_{2})$
bin is divided by the mean number of pairs in a bin
(averaged over all $(x_{1},x_{2})$ bins). This two-dimensional plot is
uniform when no inter-particle correlations are present in the
system. Correlations due to the Bose statistics
produce a ridge along the diagonal of the $(x_{1},x_{2})$ plot, which 
starts at $(0,0)$ and ends at $(1,1)$, whereas temperature 
fluctuations lead to a saddle shaped structure \cite{Tra00}.

\indent
As will be seen in the figures, the distribution of $x_{1}$ or $x_{2}$
obtained from the two-dimensional $(x_{1},x_{2})$ plots by projecting on
the $x_{1}$ or $x_{2}$ axis is not flat. This is due to the
method by which the plots are constructed. Namely, each pair of particles 
is represented by a point on the plot. Therefore, the events with higher 
multiplicities are represented by a larger number of pairs than those with 
smaller multiplicities. Since the shape of the $p_T$ distribution depends 
on the event multiplicity, the projection of the two-dimensional plot on 
$x_{1}$ or $x_{2}$ is no longer flat. However, it should 
be stressed that in the absence of any correlations the 
$(x_{1},x_{2})$ plot is uniformly populated and the $x_{1}$ and $x_{2}$  
projections are flat. 

\section{Experimental Set-up}
\label{s:experiment}

\begin{figure}[h]
\begin{center}
\epsfig{file= 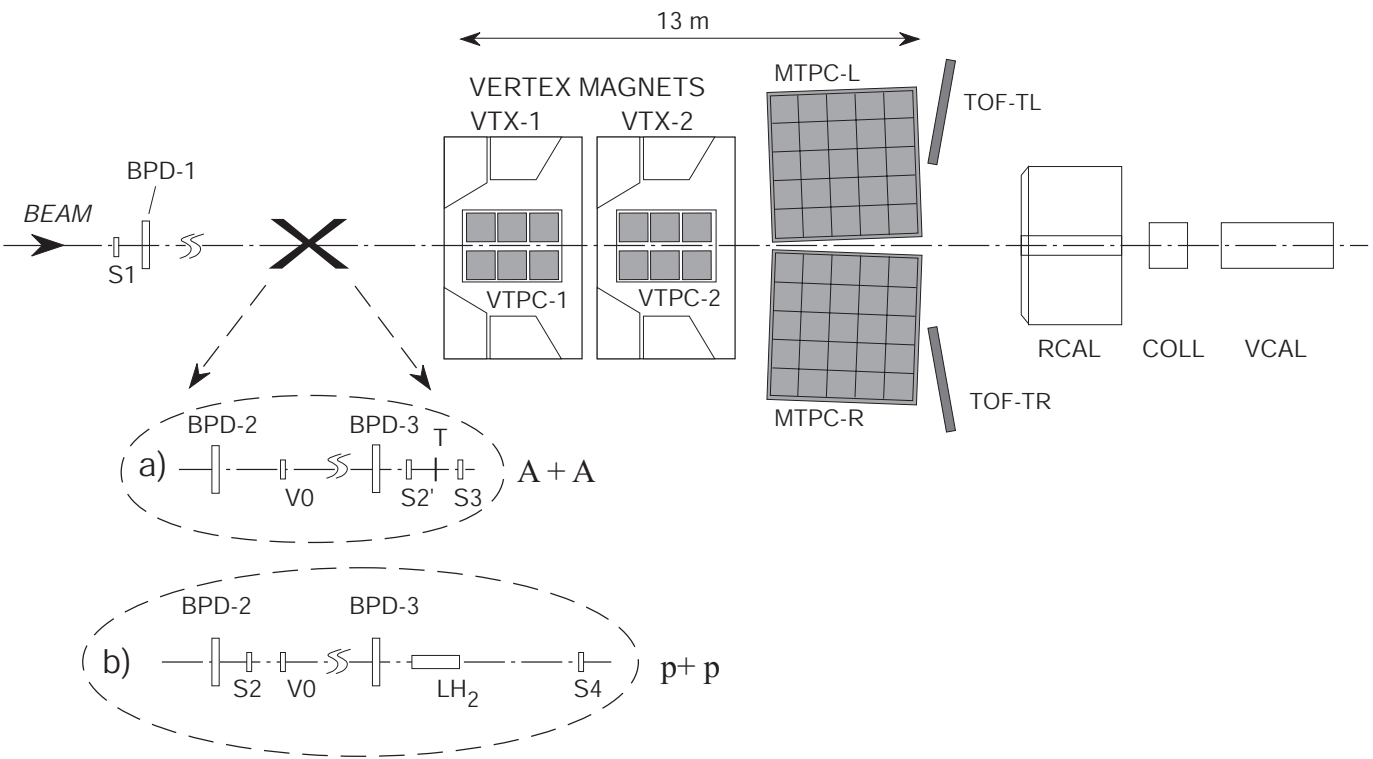,width=10cm}
\end{center}
\caption{The experimental set--up of the NA49 experiment
\cite{na49_nim} with different beam definitions and target arrangements.}
\label{setup}
\end{figure}

The NA49 experiment is a large acceptance hadron spectrometer at the CERN-SPS
used to study the hadronic final states produced by collisions of 
various beam particles (p, Pb from the SPS and C, Si from
the fragmentation of the primary Pb beam) with a variety 
of fixed targets. The main tracking devices are four large volume 
Time Projection Chambers (TPCs) (Fig. \ref{setup}) which are capable of
detecting 80\% of approximately 1500 charged particles created in a central Pb+Pb 
collision at 158 $A$GeV. Two of them, the Vertex TPCs (VTPC-1 and VTPC-2), are
located in the
magnetic field of two super-conducting dipole magnets (1.5 and 1.1 T, 
respectively) and two others (MTPC-L and MTPC-R) are positioned 
downstream of the magnets symmetrically to the beam line. The results presented
here are analyzed with a global tracking scheme \cite{na49_global}, which
combines track segments that belong to the same physical particle but were 
detected in different TPCs. The NA49 TPCs allow precise measurements
of particle momenta $p$ with a resolution 
of $\sigma(p)/p^2 \cong (0.3-7)\cdot10^{-4}$ (GeV/c)$^{-1}$.  
The set--up is supplemented by two Time of Flight (TOF) detector arrays and 
a set of calorimeters.

\indent
The targets, C (561 mg/cm$^{2}$), Si (1170 mg/cm$^{2}$) and Pb (224 mg/cm$^{2}$)
for ion collisions and a liquid hydrogen cylinder (length 20
cm) for elementary interactions, are positioned about 80 cm upstream
from VTPC-1. 

\indent
Pb beam particles are identified by means of their
charge as seen by a Helium Gas-Cherenkov counter (S2') and p beam
particles by a 2 mm scintillator (S2). Both of these are situated in
front of the target. The study of C+C and Si+Si reactions is
possible through the generation of a secondary fragmentation beam
which is produced by a primary target (1 cm carbon) in the extracted 
Pb-beam. With the proper setting of the beam line magnets
a large fraction of all $Z/A = 1/2$ fragments at $\approx 158 A$GeV
are transported to the NA49 experiment. On-line selection based
on a pulse height measurement in a scintillator beam counter (S2) is
used to select particles with $Z=6$ (Carbon) and $Z=13,14,15$ (Al, Si, P).
In addition, a measurement of the energy loss in beam position
detectors (BPD-1/2/3 in Fig. \ref{setup}) allows for a further selection 
in the off-line analysis. These detectors consist of pairs of proportional 
chambers and are placed along the beam line. They also provide a precise 
measurement of the transverse positions of the incoming beam particles. 

\indent
For p, C and Si beams, interactions in the target are selected by an
anti-coincidence of the incoming beam particle with a small scintillation
counter (S4) placed at the beam axis between the two vertex magnets. For
p+p interactions at 158 $A$GeV this counter selects a (trigger) cross
section of 28.5 mb out of 31.6 mb of the total inelastic cross section. 
For Pb beams, an interaction trigger is provided by an anti-coincidence 
with a Helium Gas-Cherenkov counter (S3)
directly behind the target. The S3 counter is used to select minimum
bias collisions by requiring a reduction of the Cherenkov signal by a
factor of about 6. Since the Cherenkov signal is proportional to
$Z^2$, this requirement ensures that the Pb projectile has interacted
with a minimal constraint on the type of interaction. This setup
limits the triggers on non-target interactions to rare beam-gas
collisions, the fraction of which proved to be small after cuts, even
in the case of peripheral Pb+Pb collisions. 

\indent
The centrality of the nuclear collisions is selected by use of  
information from a downstream calorimeter (VCAL), which measures the
energy of the projectile spectator nucleons. The geometrical acceptance of
the VCAL calorimeter is adjusted in order to cover the projectile
spectator region by the setting of the collimator (COLL). 

Details of the NA49 detector set-up and performance of tracking software are
described in \cite{na49_nim}.

\section{Data selection and analysis}
\label{s:data}

\subsection{Data sets}
The data used for the analysis consists of samples of p+p,
C+C, Si+Si and Pb+Pb collisions at 158 $A$GeV. For Pb+Pb interactions a
minimum bias trigger was used allowing a study of the centrality
dependence. The distribution of energy measured in the VCAL 
for the minimum bias Pb+Pb events was divided into six centrality 
bins (Table \ref{data_sets}) which are numbered from 1 (the most 
central) to 6 (the most peripheral). For each bin of centrality the range 
of the impact parameter $b$, and the mean number of wounded nucleons
$\langle N_{W} \rangle$ were determined by use of the Glauber model
and the VENUS event generator \cite{venus_ref}. The fraction of the total 
inelastic cross section of nucleus+nucleus collisions ($\sigma/\sigma_{tot}$)
corresponding to each data set was calculated directly by use of the 
distribution of energy measured in the VCAL. In order to estimate the 
correlation between the energy deposited in the VCAL and the impact 
parameter minimum bias VENUS events were processed through the GEANT 
detector simulation code, and the energy deposited in the VCAL was 
simulated. The correlation between $b$ and $\langle N_{W} \rangle $ was 
obtained from Glauber model calculations using the spectator-participant 
model of A+A interactions. The values of $\sigma/\sigma_{tot}$,
$ \langle N_{W} \rangle$ and $b$ presented in Table \ref{data_sets} are 
taken from \cite{Glen} for minimum bias Pb+Pb collisions and from 
\cite{Claudia} for C+C and Si+Si interactions.

\indent
The minimum bias Pb+Pb data consists of data taken during three different 
periods and at both magnetic field polarities. The most central Pb+Pb 
events correspond to 5$\%$ of the total geometric cross section. Since 
the minimum bias data provide only a small number of events in the most 
central Pb+Pb bin, additional central trigger runs were used.  

\subsection{Event and particle selection}
The aim of the event selection criteria is to reduce a possible
contamination with non-target collisions. The primary vertex was
reconstructed by fitting the intersection point of the measured
particle trajectories. Only events with a proper quality and position
of the reconstructed vertex are accepted in this analysis. The 
vertex coordinate $z$ along the beam has to satisfy
$|z-z_{0}|<\Delta z$, where the nominal vertex position $z_{0}$
and cut parameter $\Delta z$ values are: -579.5 and 5.5 cm, -579.5 and 1.5 cm, -579.5 
and 0.8 cm, -578.9 and 0.4 cm for p+p, C+C, Si+Si and Pb+Pb collisions, respectively.  

\indent
In order to reduce the contamination of particles from secondary
interactions, weak decays and other sources of non-vertex tracks, several track
cuts are applied. The accepted particles are required to have
measured points in at least one of the Vertex TPCs. 
A cut on the so-called track impact parameter, the distance between the
reconstructed main vertex and the track in the target plane, is applied 
($|b_x|<2$ cm and $|b_y|<1$ cm) to reduce the contribution of non-vertex 
particles. Moreover, particles are accepted only when the potential number 
of points (calculated on the basis of the
geometry of the track) in the detector exceeded 30. The ratio 
of the number of points on a track to the potential number of points
is required to be higher than 0.5 in order to avoid the counting of track 
segments instead of whole tracks. Only forward rapidity tracks (4.0 $< 
y_{\pi} <$ 5.5, rapidity calculated assuming the pion mass for all 
particles) with 0.005 $< p_{T} <$ 1.5 GeV/c are used in this analysis. 

\begin{figure}[h]
\begin{center}
\epsfig{file= 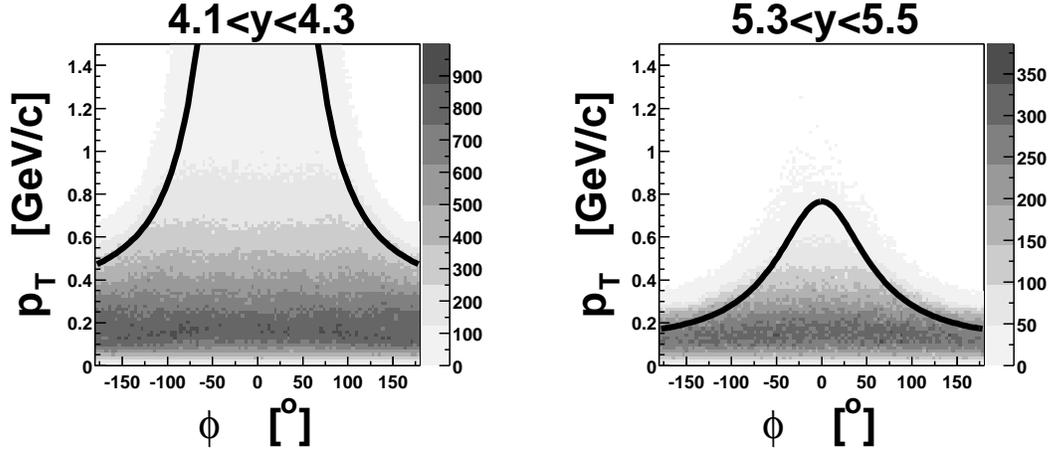,width=15cm}
\end{center}
\caption{NA49 $\phi-p_{T}$ acceptance of positively charged particles 
(standard configuration of magnetic field) for two selected rapidity bins 
at 158 $A$GeV. The solid lines represent the analytical parametrization.}
\label{angle}
\end{figure}

\indent
The NA49 detector provides a large (but not complete) acceptance in the 
forward hemisphere. Two example plots of $p_{T}$ versus azimuthal angle
$\phi$ (see Fig. \ref{setup} for definition) for positively charged 
particles (for the standard polarity of the magnetic field) are shown in 
Fig \ref{angle}. The solid lines represent a parametrization of the 
acceptance limits by the formula:
\begin{equation}
p_{T}(\phi)=\frac{1}{A+\phi^{2}/C}+B,
\label{accept}
\end{equation}
where the values of parameters $A$, $B$ and $C$ depend on the rapidity 
interval as given in Table \ref{abc}. These values apply to negatively 
charged particles as well, provided $\phi$ in Eq. \ref{accept} is 
replaced by $\phi/|\phi|(180-|\phi|)$. Only particles within the 
analytical curves are used in this analysis. This well defined acceptance 
is essential for later comparison of the results with models and other 
experiments.       

\indent
The NA49 detector is able to register particles produced
in a significantly wider rapidity range covering almost the whole forward
hemisphere. It would be interesting to determine the correlation measure
$\Phi_{p_{T}}$ in the mid-rapidity region. Preliminary results for the
rapidity range  2.9 - 4.0 were reported \cite{Volker}.
However, the azimuthal acceptance in this region is more limited and
systematic uncertainties affecting $\Phi_{p_{T}}$ are not yet fully
understood.

\subsection{Corrections and error estimates}
\label{s:corrections}

\indent
The statistical error on $\Phi_{p_{T}}$ was estimated as follows. The
whole sample of events was divided into 30 subsamples. The value of
$\Phi_{p_{T}}$ was evaluated for each subsample and the dispersion ($D$)
of the results was then calculated. The statistical error of
$\Phi_{p_{T}}$ was taken to be equal to $D/\sqrt{30}$.

\indent
The event and track selection criteria reduce the possible systematic
bias of the measured $\Phi_{p_{T}}$ values. In order to estimate the
remaining systematic uncertainty, the values of cut parameters have been varied
within a reasonable range and the systematic error has been estimated as 
a half of the difference between the highest and the lowest $\Phi_{p_{T}}$ 
value. In addition, results obtained from the analysis of data taken at 
two different magnetic field polarities as well as from different running 
periods have been compared.

\begin{figure}[h]
\begin{center}
\epsfig{file= 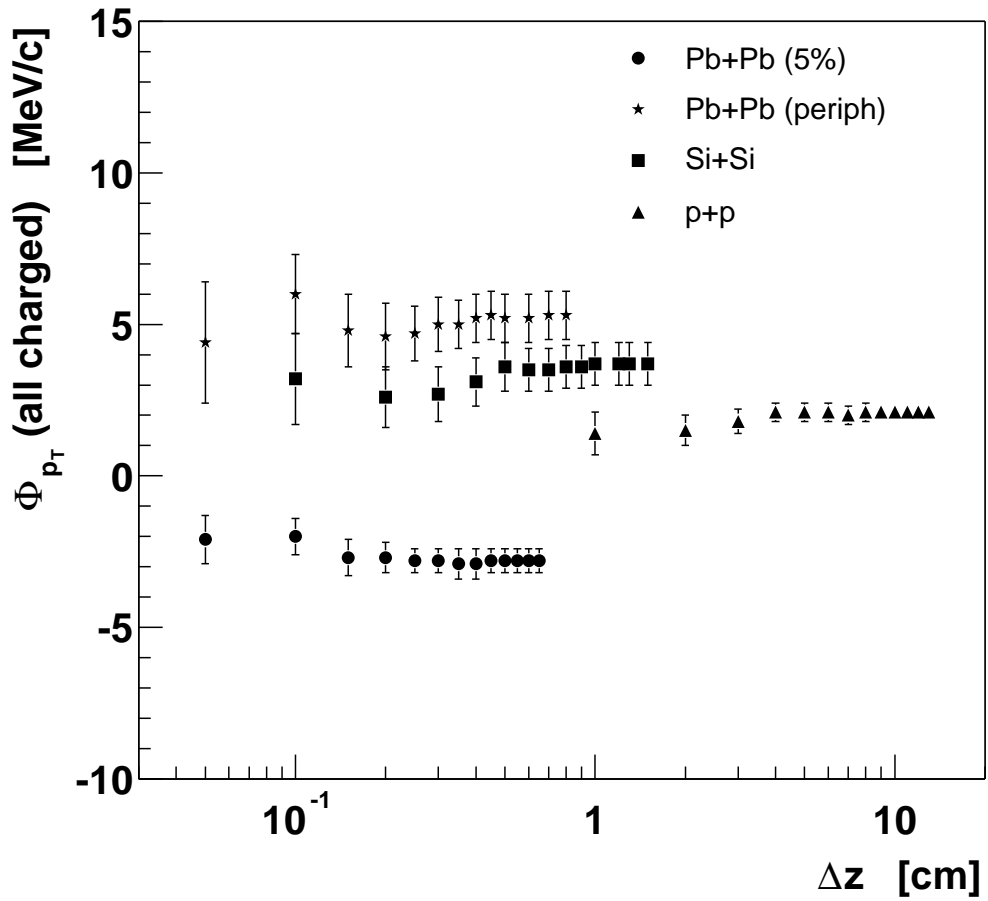,width=15cm}
\end{center}
\caption{The dependence of $\Phi_{p_{T}}$ on the allowed distance  
$\Delta z$ of the reconstructed event vertex from its nominal position.  
Note: the values and their errors are correlated.}
\label{fipt_vz}
\end{figure}

\indent
Event cuts are used to reject possible contamination of
non-target interactions, however there is always a small fraction of
remaining non-target events which can influence the $\Phi_{p_{T}}$ values.
The dependence of $\Phi_{p_{T}}$ on the event selection cut $\Delta z$ is shown in Fig.
\ref{fipt_vz}. The observed variation of $\Phi_{p_{T}}$ with $\Delta z$ is small.  
The estimated systematic error is smaller than 0.85 MeV/c 
for peripheral Pb+Pb collisions, 0.55 MeV/c for Si+Si data and 0.5 MeV/c 
for p+p events. 

\begin{figure}[h]
\begin{center}
\epsfig{file= 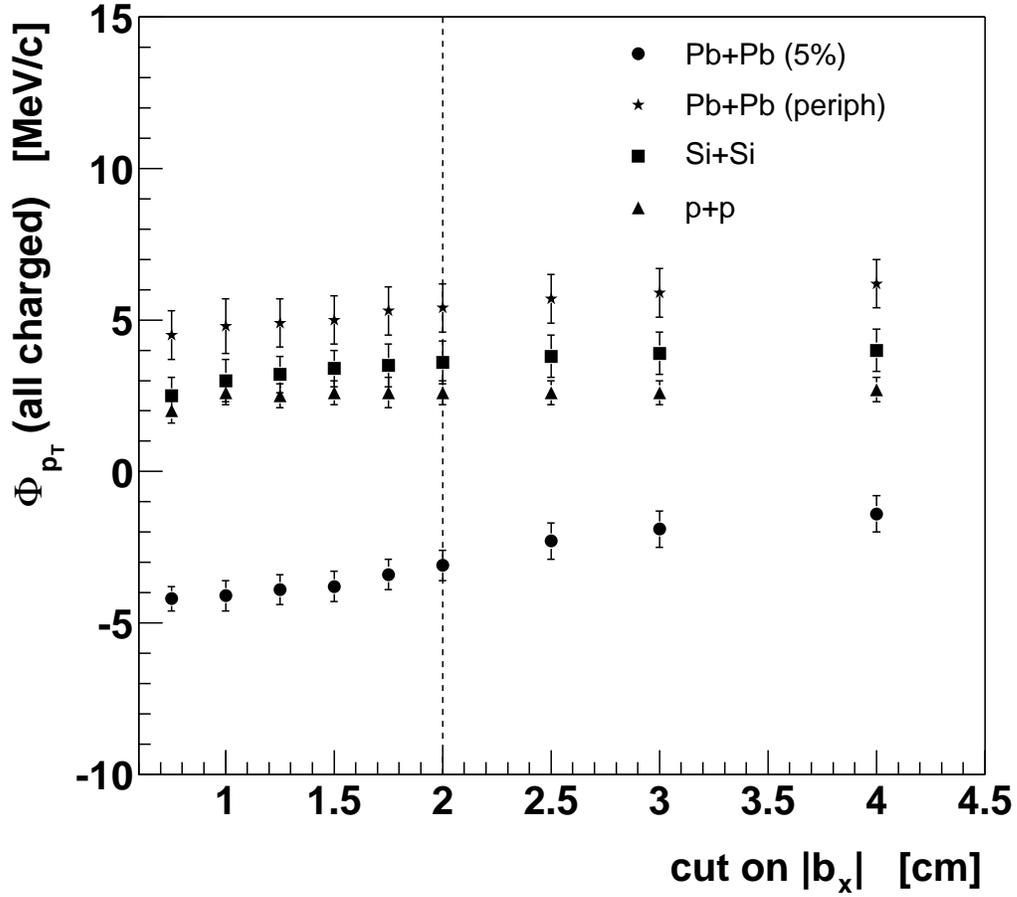,width=15cm}
\end{center}
\caption{The dependence of $\Phi_{p_{T}}$ on the upper cut in the impact 
parameter $|b_x|$. For each point the cut on $|b_y|$ was equal half the 
cut on $|b_x|$. Note: the values and their errors are correlated. The dashed 
line indicates the cut used in the analysis.}
\label{fipt_bxby}
\end{figure}

\indent
The majority of tracks selected by the track selection criteria are main
vertex tracks and the remaining fraction ($\approx$10\%) originates 
predominantly from weak
decays and secondary interactions with the material of the detector. In order to
estimate the influence of this contamination on the measured value of $\Phi_{p_{T}}$,
the impact parameter cut was varied (Fig. \ref{fipt_bxby}). 
A small increase of $\Phi_{p_{T}}$ with increasing impact parameter cut 
is observed and may be due to the increased contribution of non-vertex 
tracks from weak decays and secondary interactions. By use of the VENUS 
\cite{venus_ref} simulations these tracks were found to be correlated in 
$p_{T}$ thus increasing the measured value of 
$\Phi_{p_{T}}$. The estimated systematic error due to the contamination of 
non-vertex tracks is smaller than 1.6 MeV/c for central Pb+Pb collisions, 
0.75 MeV/c for Si+Si data and 0.35 MeV/c for p+p events. 

\begin{figure}[h]
\begin{center}
\epsfig{file= 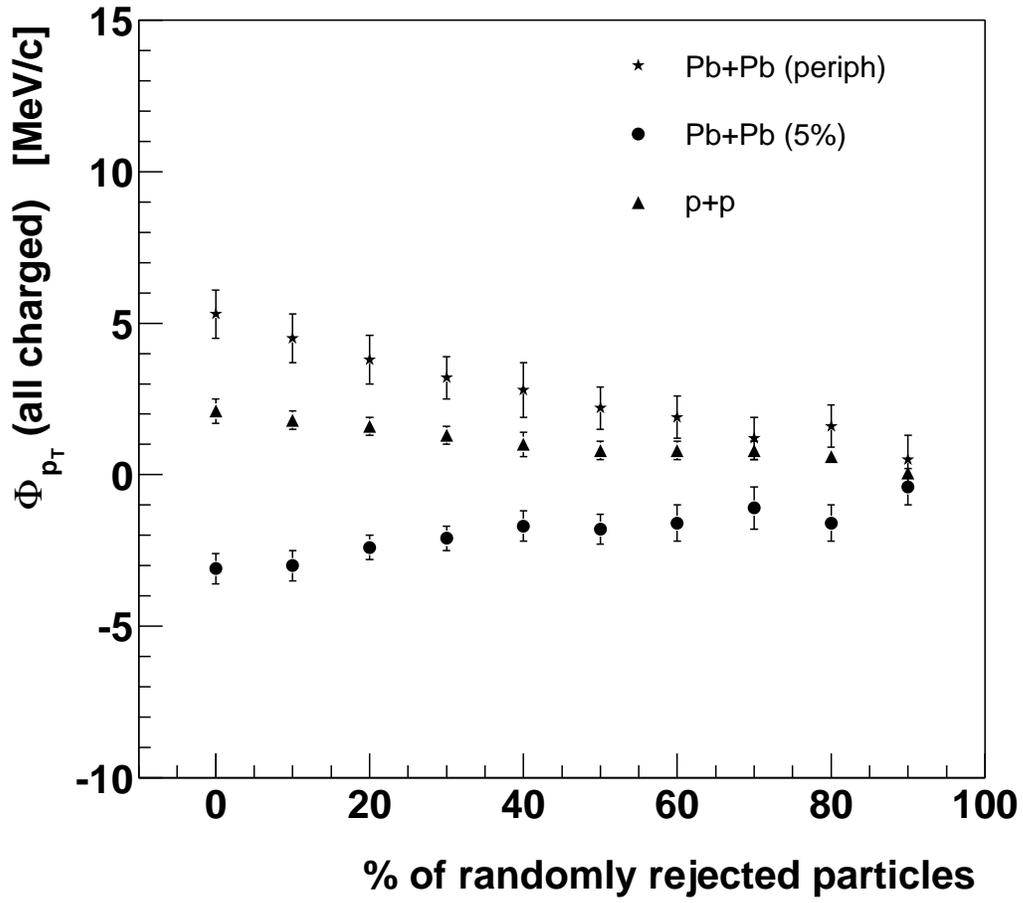,width=15cm}
\end{center}
\caption{The dependence of $\Phi_{p_{T}}$ on the fraction of 
randomly rejected particles.}
\label{random}
\end{figure}

\indent
Losses of tracks due to the reconstruction inefficiency and track selection cuts
influence the measured $\Phi_{p_{T}}$ values. In order to estimate
this effect, the dependence of $\Phi_{p_{T}}$ on the percentage
of randomly rejected particles was calculated. These dependences for 
the most peripheral (6), the most central (1) Pb+Pb collisions and for p+p 
interactions are shown in Fig. \ref{random}. Within the considered 
kinematic region (forward rapidity) the tracking efficiency of our 
detector is higher than 95\%. Fig. \ref{random} implies that the bias due 
to tracking inefficiency is not higher than 0.5-1.0 MeV/c.   

\indent
As an estimate of the systematic error on $\Phi_{p_{T}}$ a maximal error 
resulting from the above study has been taken. The systematic error is 
about 1.6 MeV/c for Pb+Pb collisions and 1.2 MeV/c for p+p, C+C and Si+Si 
interactions.   

\begin{figure}[h]
\begin{center}
\epsfig{file= 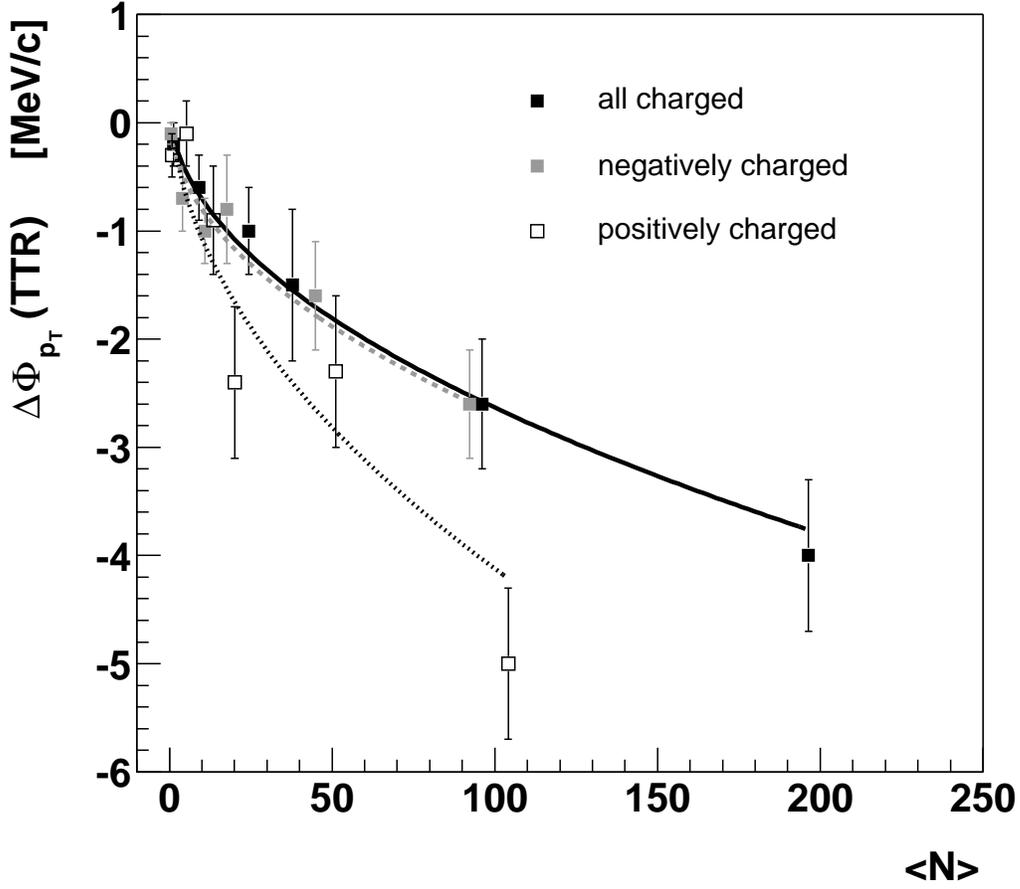,width=15cm}
\end{center}
\caption{Additive correction  $\Delta\Phi_{p_{T}}$ for limited two track
resolution effect versus multiplicity $\langle N \rangle$ of accepted 
particles.
Different points correspond to positively charged, negatively charged and 
all charged particles. The presented corrections have been determined 
using p+p, C+C, Si+Si and three centralities of Pb+Pb collisions at 158 
$A$GeV. The lines represent the analytical parametrization: 
$\Delta \Phi_{p_{T}} (\langle N \rangle)=-a\sqrt{\langle N \rangle}+b$ 
with $a$ and $b$ being parameters of a fit to the data points.}
\label{ttr}
\end{figure}

\indent
It has already been shown \cite{App99} that the
limited two track resolution influences the measured $\Phi_{p_{T}}$
values. In order to estimate this contribution several samples of mixed
events (for different A+A collisions) were produced. Mixed events 
were constructed from original events, the multiplicities of mixed 
events being the same as in the case of real events but each particle in a mixed event
taken at random from a different real event. The $\Phi_{p_{T}}$ value
calculated for the sample of mixed events was consistent with zero.
In the second step the mixed events were processed by the NA49 
simulation software. The resulting simulated raw data were 
reconstructed and the $\Phi_{p_{T}}$  measure calculated. The 
obtained $\Phi_{p_{T}}$ values
are negative as expected for the anti-correlation introduced by the
losses due to the limited two track resolution. The additive two track
resolution correction is calculated as the difference 
($\Delta\Phi_{p_{T}}$) between the values of $\Phi_{p_{T}}$
after and before this procedure. Fig. \ref{ttr} presents this correction
versus mean multiplicity. The lines correspond to an analytical
parametrization of this dependence. The absolute values of the 
track resolution corrections are larger for heavier colliding systems 
where the density of tracks is
relatively high. The absolute values of $\Delta\Phi_{p_{T}}$ are
also larger for positively charged particles than for negatively
charged ones, which is mainly due to higher track density for positive
particles caused by significantly larger number of protons than
anti-protons. The $\Delta\Phi_{p_{T}}$ values are negative indicating that
$\Phi_{p_{T}}$ measured with an ideal detector would be higher.
For a given multiplicity, $\Phi_{p_{T}}$ corrected for the
limited two track resolution effect equals `raw' $\Phi_{p_{T}}$ minus the
corresponding $\Delta\Phi_{p_{T}}$.

\section{Results}
\label{s:results}

\indent
The results shown in this section refer to {\em accepted} particles, i.e.
particles that are accepted by the detector and pass all 
kinematic cuts and track selection criteria. The data cover a broad range 
in $p_{T}$ ($0.005 < p_{T} < 1.5 $ GeV/c). The rapidity of accepted
particles is restricted to the interval 4.0 to 5.5 which corresponds
to forward rapidities in the collision of equal mass nuclei
(at 158 $A$GeV energy the center of mass rapidity equals 2.9 for fixed 
target geometry), where the azimuthal acceptance is large.   

\indent
The mean multiplicities of accepted particles, the dispersions
$\sigma_N=\sqrt{\langle N^2 \rangle -\langle N \rangle ^2}$ of
the multiplicity distributions, the mean inclusive transverse momenta,
the dispersions $\sigma _{p_T}$ of inclusive
transverse momentum distributions and $\Phi_{p_{T}}$ values for
all data sets used in this analysis are given in Table \ref{data}. The
$\Phi_{p_{T}}$ values shown in this table have been calculated for
all accepted charged particles as well as for the negatively and the
positively charged particles separately. All values of $\Phi_{p_{T}}$ 
were corrected for the two track resolution effect.

\begin{figure}[h]
\begin{center}
\epsfig{file= 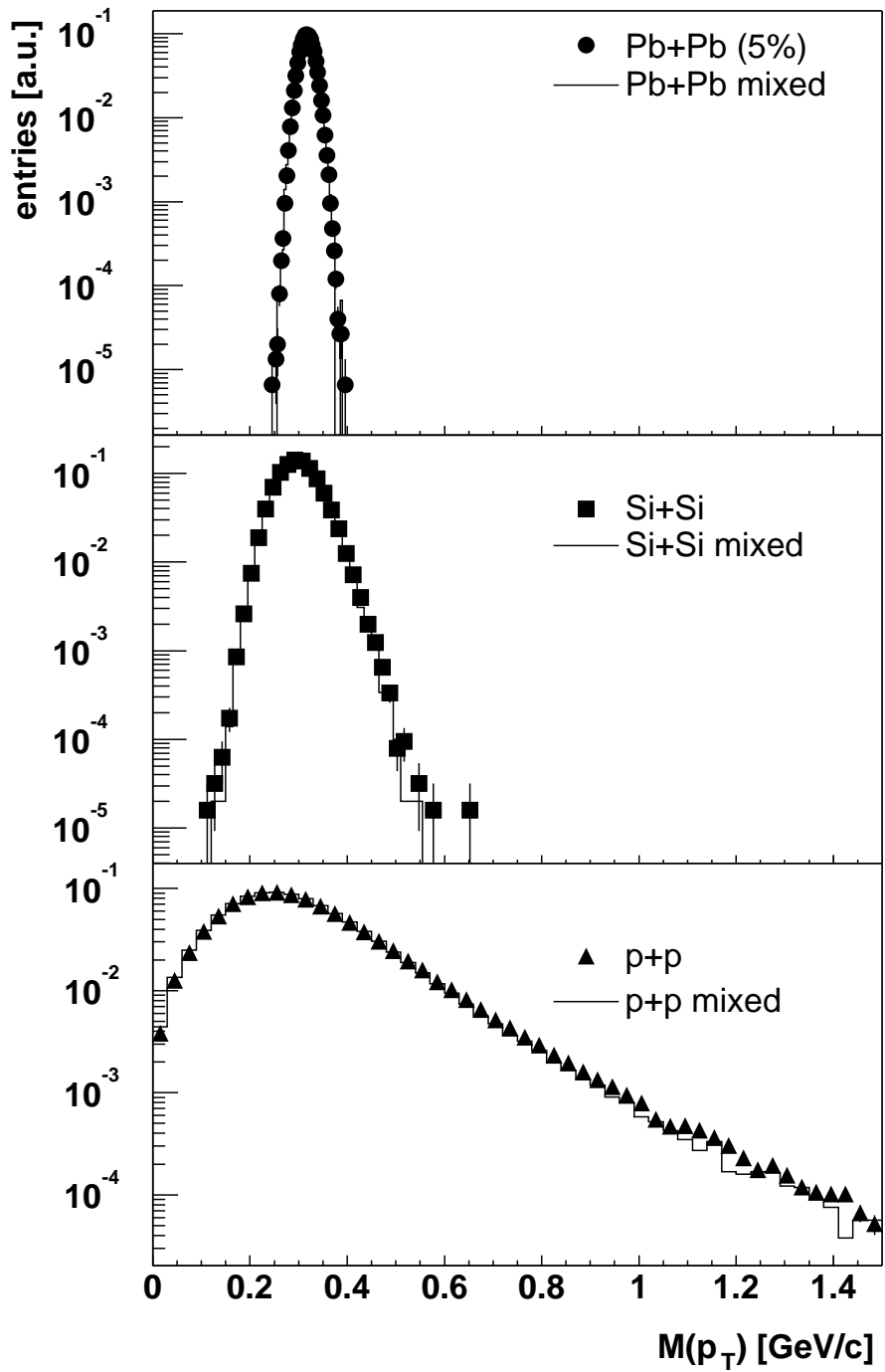,width=15cm}
\end{center}
\caption{Distributions of mean transverse momentum for real (data points) 
and mixed events (histograms). Data points are not
corrected for acceptance and limited two track resolution. Events
with accepted particle multiplicity equal to zero are not used.}
\label{meanpt}
\end{figure}

\indent
Fig. \ref{meanpt} shows the distributions of the mean (per event) 
transverse momentum $M(p_T)$ for p+p, Si+Si and central Pb+Pb collisions. 
Points correspond to data and the histograms to mixed events. The data are 
not corrected for any experimental effects. Events with zero accepted 
particle multiplicity are not taken into account. The small difference 
between $M(p_{T})$ distributions for real and mixed events demonstrates that 
dynamical fluctuations are small. Moreover, no distinct class of events 
with unusual fluctuations is observed. The width of the $M(p_{T})$ 
distribution strongly decreases with the colliding system size as expected 
from the increasing particle multiplicity. 

\begin{figure}[h]
\begin{center}
\epsfig{file= 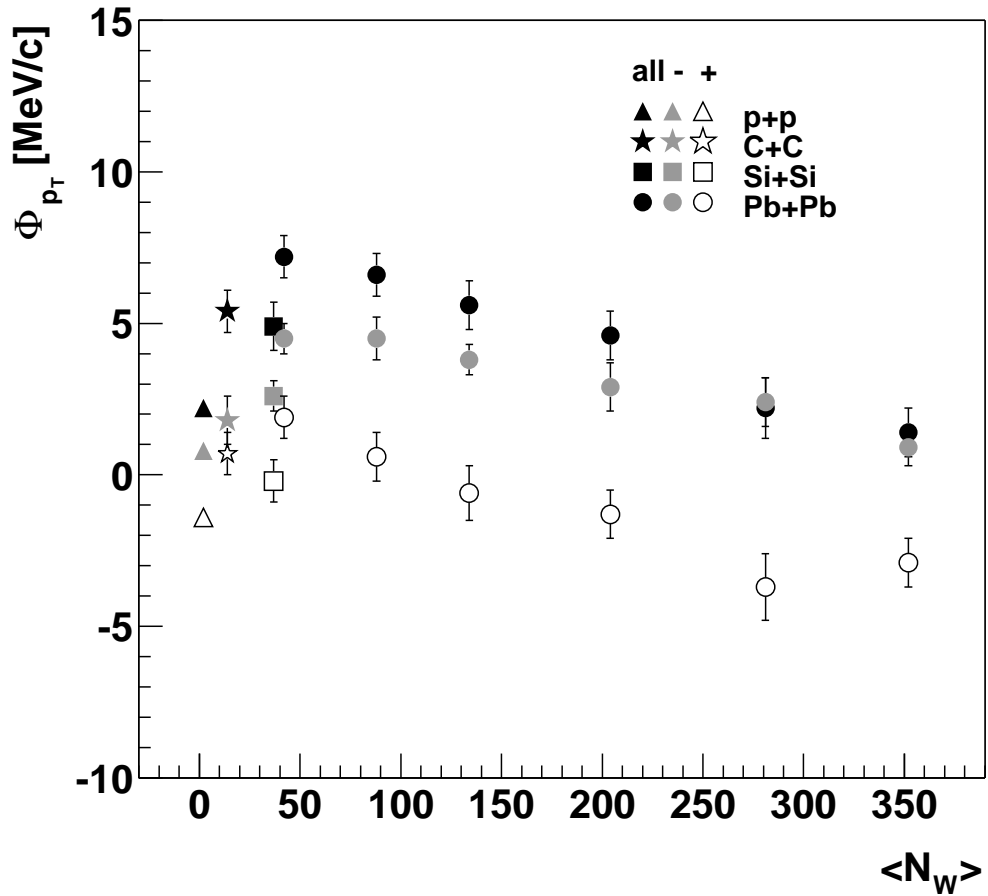,width=15cm}
\end{center}
\caption{$\Phi_{p_{T}}$ versus mean number of wounded nucleons $\langle 
N_W \rangle$. Data points were corrected for limited two track resolution.
Errors are statistical only. Systematic error is smaller than 1.6 
MeV/c.}
\label{fipt}
\end{figure}

\indent
The fluctuation measure $\Phi_{p_{T}}$ is more sensitive to small 
dynamical fluctuations. The measured values, corrected for two-track 
resolution, are plotted in Fig. \ref{fipt} versus mean number of wounded 
nucleons for all accepted charged particles and also for
positively and negatively charged particles separately. The $\Phi_{p_{T}}$
values are small (when compared to $\sigma _{p_T}$) for all investigated 
systems ($|\Phi_{p_{T}}|<10$ MeV/c), but a significant centrality 
dependence is observed. The $\Phi_{p_{T}}$ values increase with the system 
size up to the maximum value which is observed for the most peripheral 
Pb+Pb collisions, and then $\Phi_{p_{T}}$ values decrease with increasing 
number of wounded nucleons. The $\Phi_{p_{T}}$ value for Si+Si collisions 
seems to be lower than that for the most peripheral Pb+Pb 
ones although the number of wounded nucleons in both reactions is 
similar. This might suggest that $\langle N_W \rangle $ does not fully 
determine the fluctuations in A+A collisions. The $\Phi_{p_{T}}$ measure 
calculated for positively charged particles is always lower than that for
the negatively charged ones. The $\Phi_{p_{T}}$ value for all charged 
particles is always higher than that for either the negatively or the 
positively charged particles.

\indent
Two-particle correlation plots of the cumulant transverse momentum 
variable $x$ are presented in Fig. \ref{2d_plots_b} for
p+p, C+C, Si+Si and three centralities of Pb+Pb interactions (note the 
different color scales). It is seen that the plots are not uniformly 
populated. In particular, significant long range correlations of about 
40\% (the color scale varies from 0.75 to 1.6) are observed for the p+p 
data. This is rather unexpected when compared to the low
$\Phi_{p_{T}}$ value. These correlations are not seen when 
heavier colliding systems are studied. Instead, short range correlations 
become visible as an enhancement of the point density in the region close 
to the diagonal. They are most prominent for central Pb+Pb collisions and 
are consistent with the effect of Bose-Einstein statistics. For the most 
peripheral collisions the pattern seen in the two-particle correlation 
plot is different from that in the remaining systems. One observes an 
enhancement in the region close to $x_1 = 1$ and $x_2 = 1$.

\section{Discussion}
\label{s:discussion}

\indent 
In this section the results are compared with predictions of models and 
with the results of other experiments.

\indent
Fig. \ref{fipt} shows that $\Phi_{p_{T}}$ is a non-monotonic function of 
centrality with the maximum at approximately $N_W = 40$. Such a behavior 
strongly resembles the dependence of the magnitude of collective flow - directed
$(v_1)$ and elliptic $(v_2)$ - on $N_W$ \cite{flow_paper}. So, there is a 
natural suggestion that the $p_T$ fluctuations measured by  $\Phi_{p_{T}}$  
may be caused by the collective flow. This suggestion was checked by 
performing a simple Monte Carlo analysis generating events with 
independent particles, following the measured inclusive $p_T$ 
distribution. The uniform azimuthal angle distribution of the events 
was modified by the collective elliptic flow but the particles
remained independent from each other with respect to their transverse 
momenta. For such events the dynamical $p_T$ fluctuations vanish if the
azimuthal angle acceptance is complete. However, the azimuthal angle
acceptance of the NA49 detector is not flat, and consequently the azimuthal
anisotropy generates a finite value of $\Phi_{p_{T}}$, even though the 
$p_T$ of particles are independent of each other. It was found that an 
abnormally large value of $v_2=0.5$ leads to $\Phi_{p_{T}}$
as large as 17 MeV/c, but a realistic magnitude of $v_2$, which depends 
on $p_T$ (according to \cite{flow_paper} taken as $0.05 \cdot p_T$(GeV/c)), 
results in a value of $\Phi_{p_{T}}$ which is consistent with\
zero ($\Phi_{p_{T}} = 0.0 \pm 0.6$ MeV/c). Thus, one concludes that the 
effect of the azimuthal anisotropy caused by the collective flow combined with 
the incomplete azimuthal acceptance is not responsible for the observed 
dynamical $p_T$ fluctuations.

\begin{figure}[h]
\begin{center}
\epsfig{file= 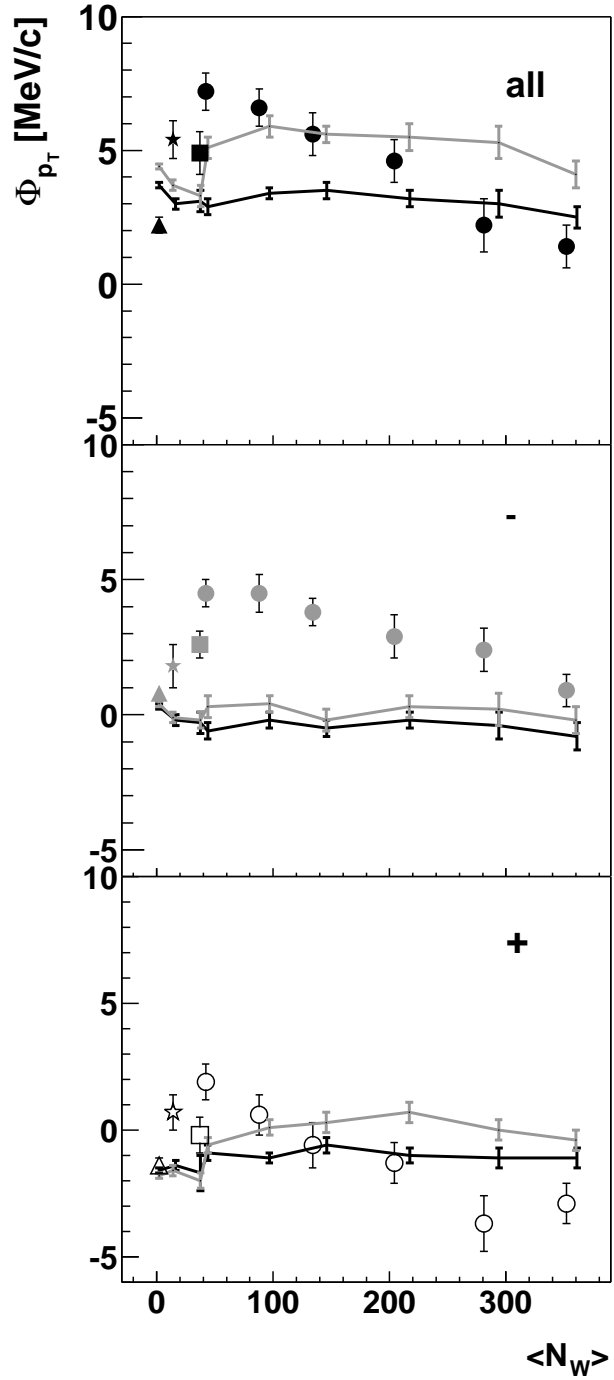,width=10cm}
\end{center}
\caption{$\Phi_{p_{T}}$ versus mean number of wounded nucleons calculated
using the HIJING model with geometrical acceptance cuts included (black 
lines) and without geometrical acceptance restrictions (gray lines).
Results are compared to data (points) corrected for
limited two track resolution (the markers are the same as in Fig.
\ref{fipt}). The panels represent: all charged,
negatively charged and positively charged particles. Data points contain 
both short and long range correlations. The effects of short range 
correlations are not incorporated in the HIJING model.}
\label{fipt_HIJING}
\end{figure}

\indent
In Fig. \ref{fipt_HIJING} the dependence of $\Phi_{p_{T}}$ on the mean 
number of wounded nucleons is directly compared to predictions of the HIJING 
\cite{HIJING_paper} model (default parameters were used) for 
all charged particles, and for negatively charged and positively charged
particles separately. The same kinematic cuts are applied as for the data. 
The black lines represent the results of the HIJING simulations where the 
effect of the limited NA49 acceptance ($p_T$ versus azimuthal angle) is taken 
into account. The gray lines refer to the HIJING predictions 
for full azimuthal acceptance to demonstrate the effect of the limited 
acceptance of the detector.  

\indent
In contrast to the data, $\Phi_{p_{T}}$ computed within the HIJING model 
does not change when going from elementary to central Pb+Pb collisions because 
the HIJING model represents an essentially independent superposition of 
N+N interactions. The effects of short range correlations (Bose-Einstein and 
Coulomb) have not been incorporated in the HIJING model. However it was 
estimated in the previous analysis \cite{App99} that the combined 
effect of short range correlations produces $\Phi_{p_{T}}$ values on the 
level of 5 MeV/c for central Pb+Pb collisions. This effect strongly 
depends on multiplicity and becomes negligible for p+p interactions. 

\indent
For the HIJING model $\Phi_{p_{T}}$ values for positively charged 
particles are, as in the case of real data, lower than for negatively 
charged and for all charged particles. The fact that $\Phi_{p_{T}}$ values 
for positively charged particles are always lower than for negatively 
charged ones has been found (using HIJING) to be related to the limited 
acceptance and treating protons as pions (assuming the pion mass for all 
produced particles). 

\indent
When $\Phi_{p_{T}}$ was first introduced \cite{Gaz92} it was believed
that its value would be non-zero for elementary interactions
(mainly due to the $M(p_T)$ versus $N$ dependence) and would
vanish for heavier colliding systems as a result of equilibration. The 
present measurements do not confirm this expectation. Although 
$\Phi_{p_{T}}$ is close to zero for central Pb+Pb collisions, 
the maximum value is observed not for p+p data, but for colliding systems 
with $N_W \simeq$ 40. 

\begin{figure}[h]
\begin{center}
\epsfig{file= 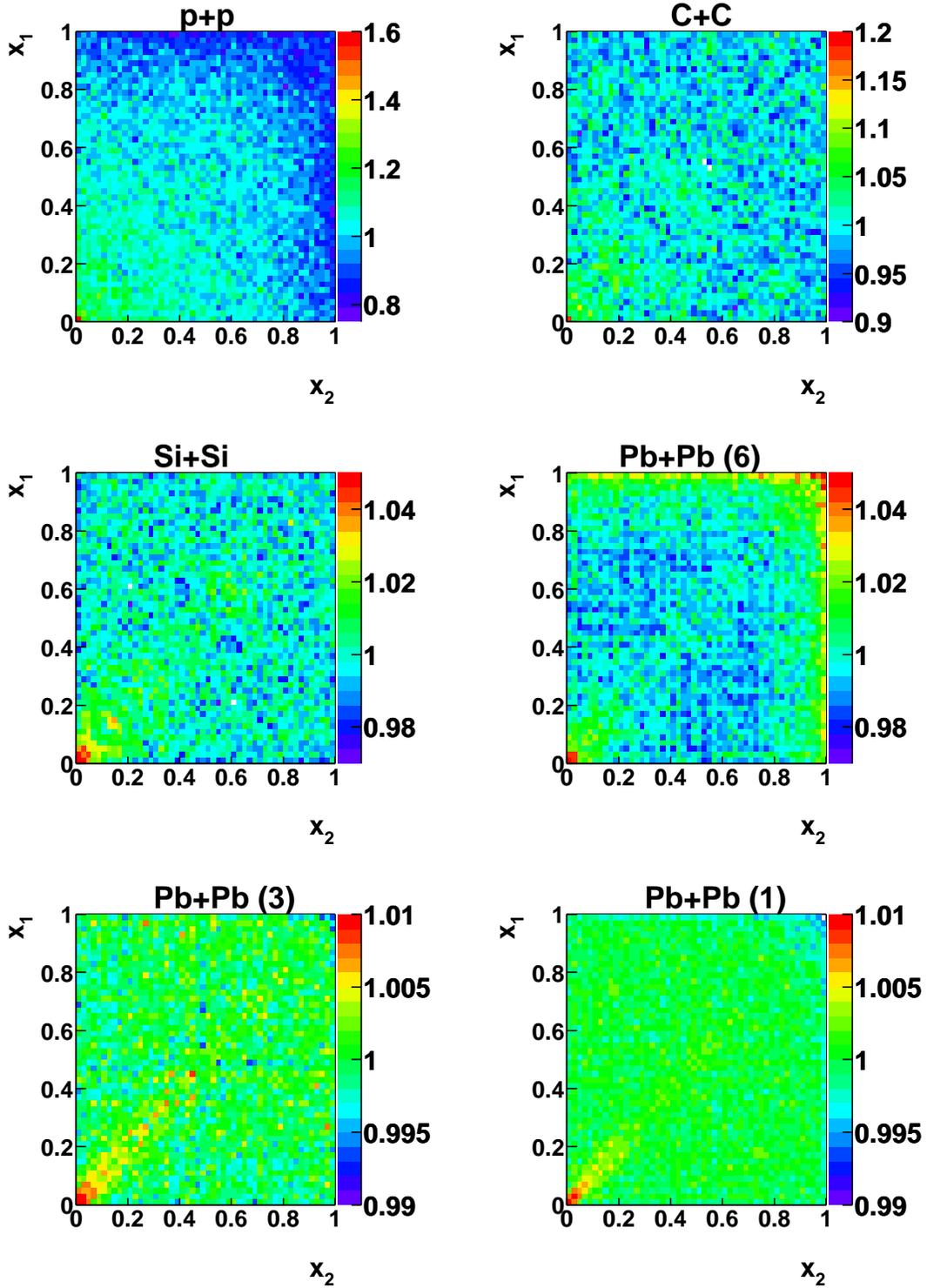,width=15cm}
\end{center}  
\caption{Two-particle correlation plots using the cumulant $p_T$ variable 
$x$. After each charged particle pair $(x_{1},x_{2})$ was entered into the 
plot, the bin contents were normalized by dividing with the average number 
of entries per bin. The data are plotted with different color scales.}
\label{2d_plots_b}
\end{figure}

\begin{figure}[h]
\begin{center}
\epsfig{file= 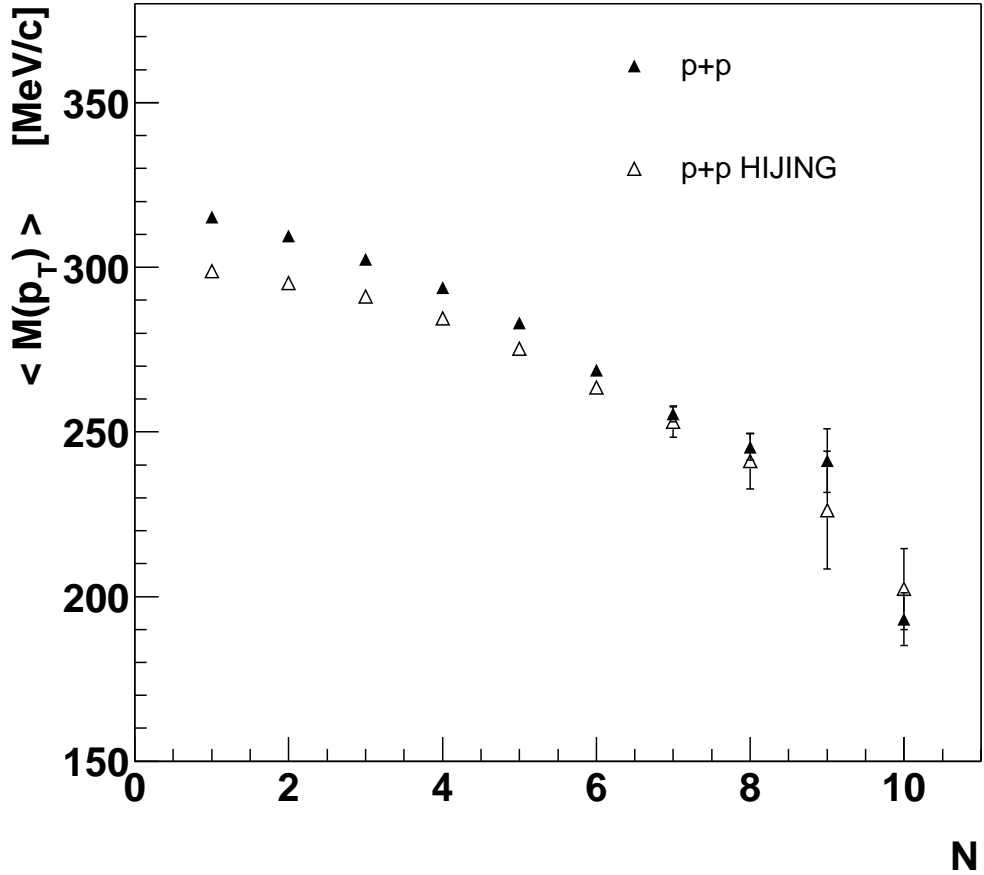,width=15cm}
\end{center}
\caption{Mean transverse momentum versus multiplicity 
of all accepted particles. The closed symbols represent p+p data at
158 $A$GeV (data are not corrected for limited two track resolution 
effect) and
the open symbols corresponds to p+p events simulated using the HIJING
model (the effects of the limited NA49 acceptance are included).
Events with accepted particle multiplicity equal zero are not used.}
\label{pt_n}
\end{figure}

\begin{figure}[h]
\begin{center}
\epsfig{file= 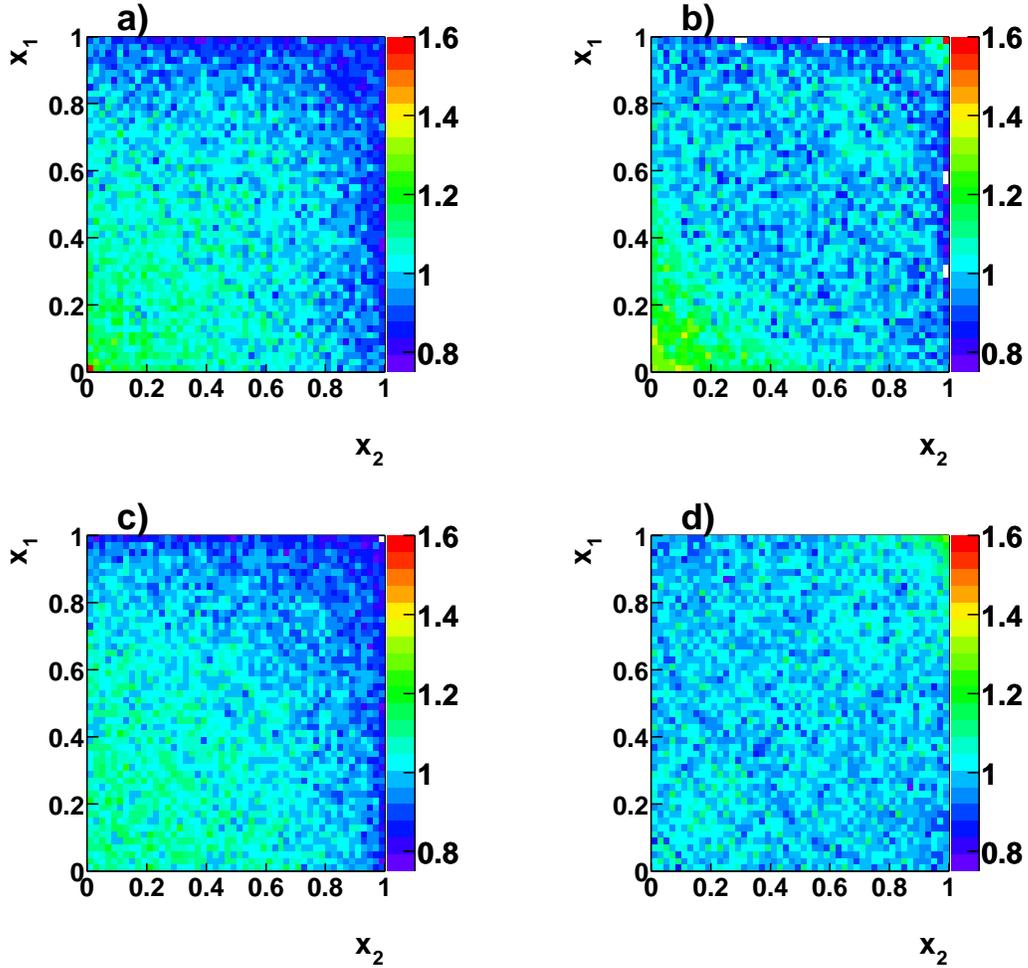,width=15cm}
\end{center}
\caption{Two-particle correlation plots using the cumulant $p_T$ variable 
$x$. Results are shown for all charged particles for p+p data (a) 
compared to: simulated p+p HIJING events with limited NA49 acceptance 
(b), simple random-generator model, which reproduces $M(p_T)$ versus $N$ 
correlation for p+p data (c), model of fluctuations of the inverse slope 
parameter for p+p data on the level of about 10 \%  (d).}
\label{pp_2D} 
\end{figure}

\indent
Although the value of $\Phi_{p_{T}}$ is small for p+p collisions, a 
significant structure appears in the two-particle correlation plot 
(Fig. \ref{2d_plots_b}). The first 
candidate for its origin is the dependence of $M(p_T)$ 
on $N$, observed for elementary interactions \cite{Gaz92}. Fig. \ref{pt_n} 
compares $\langle M(p_T) \rangle$ versus $N$ for the HIJING model and for 
real p+p data ($\langle ... \rangle$ represents averaging over events with 
a given $N$). The HIJING model includes the NA49
acceptance and all kinematic restrictions. It shows good agreement with 
the measurements. Fig. \ref{pp_2D} presents the p+p two-particle correlation 
plots for data (a), for the HIJING model (b) and for a simple random 
generator, which reproduces the dependence of $M(p_T)$ on $N$ observed in 
the data (c). The accepted particle multiplicity distribution for
the random generator (c) is the same as in the case of data (all
kinematic cuts and NA49 geometric acceptance are included). Both
models (b) and (c) qualitatively reproduce the structure of the
two-particle correlation plot observed in the data, however, 
the HIJING model (b) shows additionally a small enhancement of the point 
density in the region of high $x$.

The $\Phi_{p_{T}}$ value calculated for model (c) for all 
charged particles equals 1.2 $\pm$ 0.2 $\pm$ 1.8 MeV/c and is consistent 
with $\Phi_{p_{T}}$ for the p+p data (2.2 $\pm$ 0.3 $\pm$ 1.2 MeV/c). 
Nevertheless the 
small difference might indicate that, in agreement with the analysis 
presented in \cite{Korus:2001fv}, there could be an additional source of 
correlations present in the data. However, due to the relatively high 
statistical and systematic errors the effect cannot be estimated quantitatively.
A similar analysis was carried out for negatively charged particles
only, where the correlations caused by resonances and by charge 
conservation are expected to be smaller than for all charged particles. 
There, the experimental value of $\Phi_{p_T}$ is 0.8 $\pm$ 0.1 $\pm$ 
1.2 MeV/c while the model (c) gives 0.6 $\pm$ 0.2 $\pm$ 0.3 MeV/c. 
Two-particle correlation plots for negatively charged 
particles only are very similar for data and for model (c).  
Thus, one concludes that the results on negative particles from p+p 
interactions are consistent with the conjecture \cite{Golo03} that the 
particles are emitted independently but that their $y$ and $p_T$ 
distributions depend on event multiplicity.

\indent
The last panel (d) of Fig. \ref{pp_2D} presents the result of a simple
temperature fluctuation model (the concept is described in \cite{Kor01}), 
which assumes that the only source of fluctuations is
event-by-event fluctuation of the inverse slope parameter ($T$) of the
transverse mass spectra. The model assumes a Gaussian shaped
rapidity distribution and an exponential shape of the transverse mass
distribution with the mean inverse slope parameter $\langle T \rangle$ =
152 MeV adjusted to agree with the experimental p+p results.
All kinematic cuts applied for the real data are also used in 
this model and the effect of the finite detector acceptance is taken into
account. The mean multiplicity of all accepted particles is the same as
in the data. The fluctuations of the inverse slope parameter lead 
to a saddle shaped structure in the two-particle correlation 
plots. Panel d) shows the result for fluctuations of $T$ on the 
level of about 10\% (the dispersion $\sigma _T$ = 16 MeV). Because of the 
difference between the panels a) and d) one concludes that fluctuations of the 
inverse slope parameter are not the (main) source of correlations in p+p 
data.  

\begin{figure}[h]
\begin{center}
\epsfig{file= 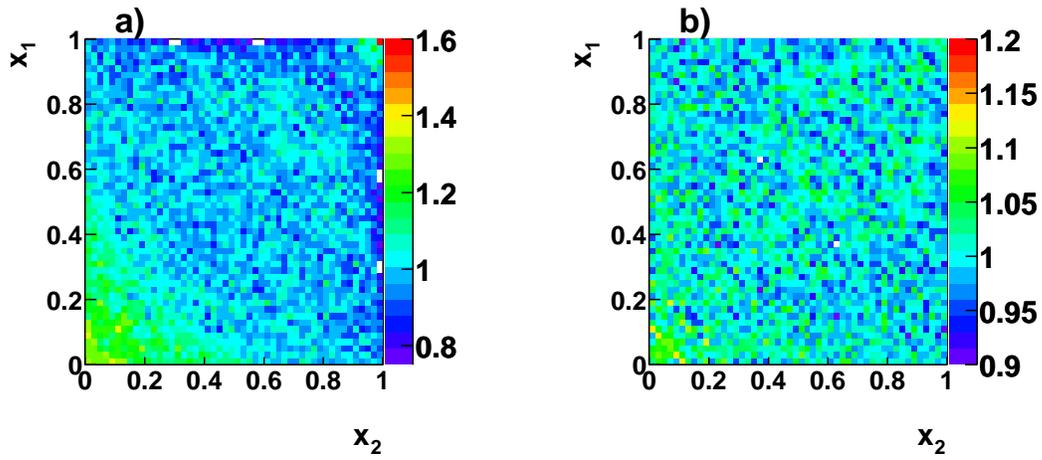,width=15cm}
\end{center}
\caption{Two-particle correlation plots using the cumulant $p_T$ variable 
$x$. Results are shown for simulated  p+p (a) and C+C (b) collisions from 
the HIJING model. Limited NA49 acceptance is taken into account.}
\label{CC_2D}
\end{figure}

\indent
The HIJING model has also been used to obtain a two-particle correlation
plot for C+C collisions (Fig. \ref{CC_2D} b), which appears to be similar 
to that observed for real events (Fig. \ref{2d_plots_b}). Fig. 
\ref{CC_2D} a (the same as Fig. \ref{pp_2D} b) presents p+p events 
simulated by the HIJING model. The structure 
observed for p+p collisions vanishes for heavier systems due to the dilution 
effect from the higher number of uncorrelated particles (resulting from 
different N+N interactions), whereas the $\Phi_{p_{T}}$ measure is not 
affected by this dilution effect.

\vspace{1cm}

\begin{figure}[h]
\begin{center}
\epsfig{file= 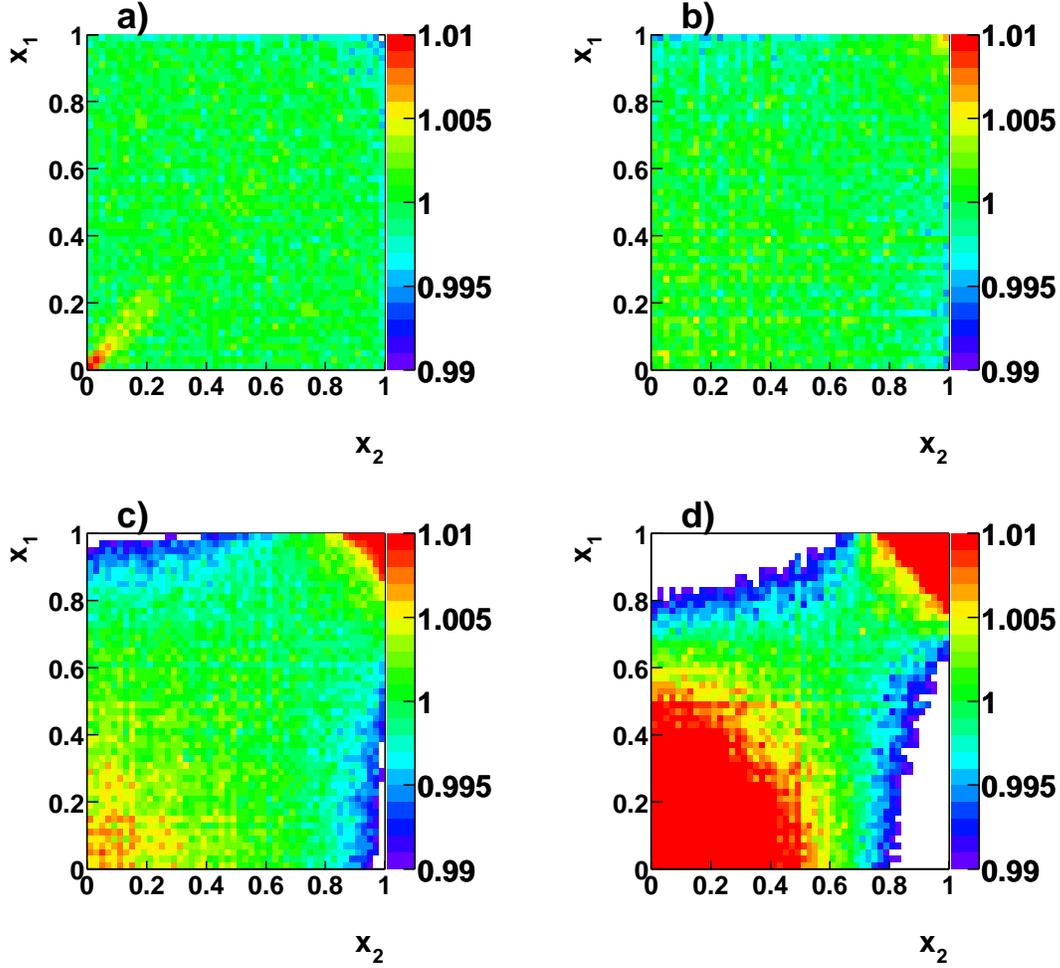,width=15cm}
\end{center}
\caption{Two-particle correlation plots using the cumulant $p_T$ variable 
$x$ for the most central Pb+Pb collisions. The experimental result is 
shown in (a) and compared to the inverse slope parameter fluctuation 
model (b-d). In the model the mean value of the inverse slope parameter 
was set to 190 MeV. The dispersions of the Gaussian shaped inverse slope 
parameter distributions were set to: $\sigma _T$ = 5 MeV (b), $\sigma _T$ 
= 10 MeV (c), $\sigma _T$ = 20 MeV (d)}
\label{toy}
\end{figure}

\begin{figure}[h]
\begin{center}
\epsfig{file= 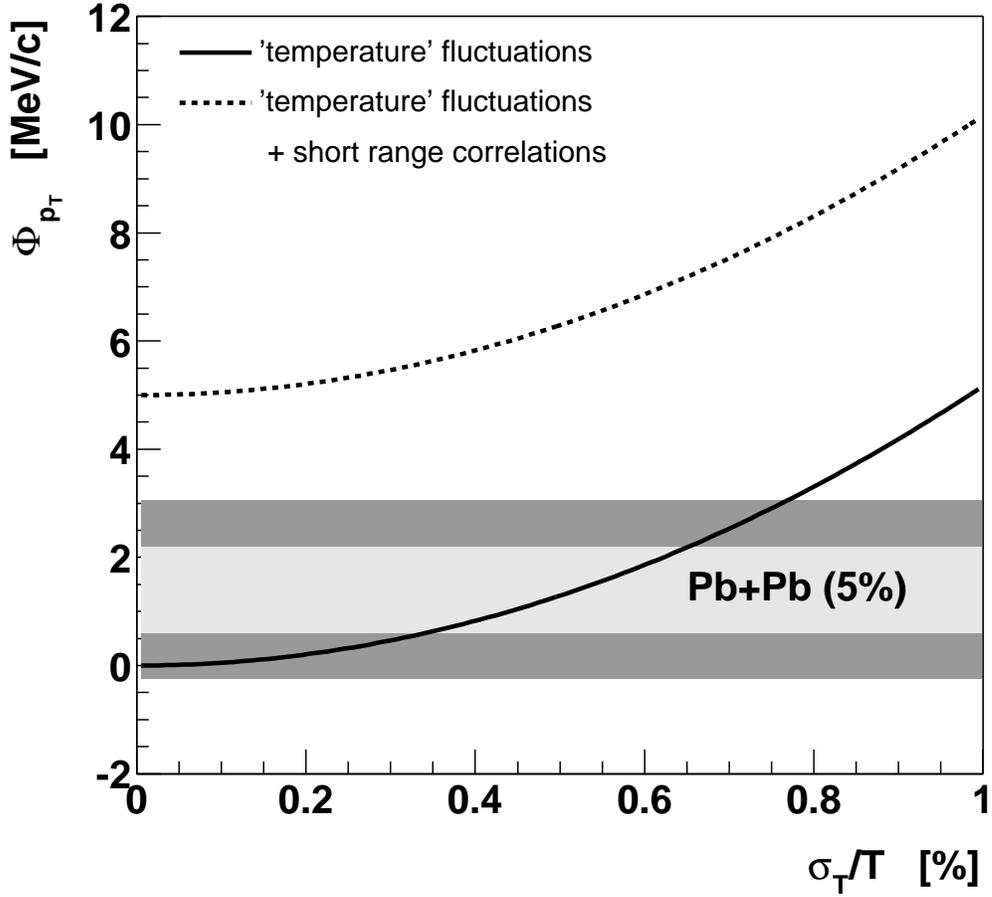,width=15cm}
\end{center}
\caption{Predicted dependence of $\Phi_{p_{T}}$ on $T$ 
fluctuations \cite{Kor01} compared to the measured $\Phi_{p_{T}}$ for  
the 5\% most central Pb+Pb collisions. The uncertainty of the measured 
$\Phi_{p_{T}}$ value is represented by the bands for 
statistical (gray) and systematic (dark gray) errors.}
\label{fipt_deltaT}
\end{figure}

\indent
In order to see how dynamical fluctuations influence two-particle
correlation plots for central Pb+Pb data the above model with 
fluctuations of the inverse slope parameter was used again. The mean 
inverse slope parameter $\langle T \rangle$ was set to 190 MeV and 
the mean multiplicity of all accepted particles to 200 in
order to compare the results with central Pb+Pb collisions. The inverse slope 
parameter varied from event to event with a Gaussian shaped distribution 
of width $\sigma _T$. Fig. \ref{toy} presents $(x_{1},x_{2})$ plots  
for different levels of the inverse slope parameter fluctuations. The 
fluctuations lead to a saddle shaped structure which is not visible in
central Pb+Pb collisions. One can thus exclude significant $T$ 
fluctuations in central Pb+Pb collisions at top SPS energy.
In Fig. \ref{fipt_deltaT} the predicted 
dependence of $\Phi_{p_{T}}$ on $T$ fluctuations \cite{Kor01} is plotted 
and compared to $\Phi_{p_{T}}$ measured for the 5\% most central Pb+Pb 
interactions. The experimental $\Phi_{p_{T}}$ value contains both short 
and long range correlations. The solid line corresponds to the 
$T$-fluctuation model presented in \cite{Kor01} which does not include 
short range correlations. The dashed line is the
combination of this model with a contribution of short range 
correlations estimated experimentally \cite{App99} as 5 MeV/c, in 
agreement with theoretical arguments \cite{Mro98a, Mro99a}. 
One sees that the observed value of $\Phi_{p_{T}}$ is already below the 
contribution of the Bose-Einstein correlations and that the inclusion of 
slope fluctuations makes the difference even larger. Thus, one can 
conclude, in agreement with the previous results \cite{App99}, that the 
data leave no space for significant $T$ fluctuations provided they are 
not canceled by other negative correlations.

\vspace{1cm}

\indent
An increase of transverse momentum fluctuations was predicted \cite{SRS} 
to occur in A+A collisions which freeze out near a second order critical 
end point of the QCD phase diagram. Based on calculations from 
\cite{SRS} and the numbers given in Table \ref{data} it can be estimated that 
such critical fluctuations alone should result in $\Phi_{p_{T}} \simeq 20$ 
MeV/c. This number is significantly larger than the maximum value of 
$\Phi_{p_{T}}$ found in this analysis - for peripheral Pb+Pb interactions 
($\Phi_{p_{T}} = 7.2 \pm 0.7 \pm 1.6 $ MeV/c). Note, however, that in this 
theoretical estimate the effect of the limited experimental acceptance was 
not taken into account. 

\vspace{1cm}

\indent
Transverse momentum fluctuations in A+A collisions
and elementary interactions were measured by several experiments
in the SPS energy range. A value of $\Phi_{p_{T}} = 10.9 \pm 1.5 $ MeV/c 
for charged hadrons was reported by the NA22 experiment in $\pi^{+}p$ and 
$K^{+}p$ interactions \cite{NA22} at 250 GeV in a rapidity acceptance 
similar to the one used in this paper. The value obtained by the NA22 
experiment is higher than the result presented in this paper for p+p 
interactions at 158 $A$GeV ($\Phi_{p_{T}} = 2.2 \pm 0.3 \pm 1.2$ MeV/c).
This can be caused by the differences in azimuthal acceptance, energy 
and types of interacting hadrons.

\indent
The CERES experiment at the SPS measured  $\Phi_{p_{T}} \approx 5 $ MeV/c 
for charged hadrons in central Pb+Au collisions at 158 $A$GeV \cite{CERES}. 
This measurement was performed close to mid-rapidity  
(pseudo-rapidity $2.2 < \eta < 2.7$) in the $p_T$ range $0.1 < p_T < 1.5 $ 
GeV/c  and with full azimuthal angle acceptance. A somewhat smaller value, 
$\Phi_{p_{T}} \approx 1.5 $ MeV/c, is found in this paper for the similar 
Pb+Pb reaction at forward rapidities.

\indent
Significant non-statistical fluctuations ($\Phi_{p_{T}} = 52.6 \pm 0.3$  
MeV/c) were measured for charged hadrons at mid-rapidity in central
Au+Au collisions at $\sqrt{s_{NN}}$ = 130 GeV by the STAR experiment 
\cite{star_2003}. At the RHIC energies $p_T$ fluctuations are expected to 
be larger than at the SPS due to a significant contribution of correlated
particles originating from (mini-)jet fragmentation.

\indent
A non-monotonic centrality dependence of $p_T$ fluctuations,
with a maximum for semi-central Au+Au collisions, was reported by
the PHENIX experiment at $\sqrt{s_{NN}}$ = 200 GeV \cite{PHENIX2}. 
This result is in qualitative agreement with the result presented in 
this paper for Pb+Pb collisions at 158 $A$GeV. A centrality 
dependence of $p_T$ fluctuations was observed also by the STAR experiment 
at $\sqrt{s_{NN}}$ = 130 GeV \cite{star_2003}.

\vspace{1cm}

\indent
It should be stressed that currently available event-by-event measurements 
were performed in a limited acceptance, the multiplicity of 
observed particles being below 20\% of the total multiplicity. 
Consequently, the sensitivity of these measurements, in particular to 
long range correlations is reduced. It is desirable to perform 
future event-by-event measurements in an extended acceptance.

\section{Summary}

Transverse momentum event-by-event fluctuations were studied
for p+p, C+C, Si+Si and Pb+Pb collisions at 158 $A$GeV. The analysis was
limited to the forward rapidity region. Three different
characteristics were measured: the fluctuations of average transverse 
momentum ($M(p_T)$) of the event, the $\Phi_{p_{T}}$ fluctuation measure, 
and transverse momentum two-particle correlations. All 
measured $\Phi_{p_{T}}$ values are below 10 MeV/c and are much smaller than the 
dispersion of the inclusive $p_T$ distribution. However, the correlations 
observed in p+p collisions are not simply more and more diluted when going 
to heavier colliding systems as could be expected if the 
created matter approaches a higher level of equilibrium with  
increasing system size. Instead, a significant system size 
dependence of the $\Phi_{p_{T}}$ measure is seen with a maximum for 
peripheral Pb+Pb collisions with $N_W \simeq$ 
40. The two-particle correlation plot for p+p data shows a
prominent structure which was found to be connected with the 
dependence of $M(p_T)$ on $N$. This structure disappears when going to 
heavier colliding systems. Instead short range correlations 
become visible as an enhancement of the point density in the region close 
to the diagonal. This effect is strongest for the most central 
Pb+Pb interactions. No structure characteristic of event-by-event 
temperature fluctuations is observed. 

The HIJING model qualitatively reproduces 
the structure of two-particle correlation plots for p+p and C+C data.
However, in contrast to the data, it shows no centrality dependence of 
$\Phi_{p_{T}}$.  

\indent
In future, a study of the energy dependence of transverse momentum 
fluctuations in the CERN SPS energy range is planned using the NA49 Pb+Pb 
collision data taken at different beam energies. The aim is to search for 
possible anomalies connected with the onset of the deconfinement phase
transition, which is indicated by features of pion and strangeness 
production at low SPS energies \cite{na49_energy}.

\newpage
\vspace{1cm}
\noindent
{\bf Acknowledgments}

This work was supported by the Director, Office of Energy Rese
arch, 
Division of Nuclear Physics of the Office of High Energy and Nuclear 
Physics 
of the US Department of Energy (DE-ACO3-76SFOOO98 and DE-FG02-91ER40609), 
the US National Science Foundation, 
the Bundesministerium f\"ur Bildung und Forschung, Germany, 
the Alexander von Humboldt Foundation, 
the UK Engineering and Physical Sciences Research Council, 
the Polish State Committee for Scientific Research (2 P03B 130 23, 
SPB/CERN/P-03
/Dz 446/2002-2004, 2 P03B 02418, 2 P03B 04123), 
the Hungarian Scientific Research Foundation (T032648, T14920 and T32293),
Hungarian National Science Foundation, OTKA, (F034707),
the EC Marie Curie Foundation,
and the Polish-German Foundation.

\newpage

\begin{table}[h]
\begin{center}
\begin{tabular}{|c|c|c|c|c|c|}
\hline
 & No. of events & $\sigma/\sigma_{tot}$ in each bin & 
 $\langle N_{W} \rangle$ & $b$ range [fm]\cr
\hline
\hline
p+p & 570 000 & 0.9 &  2 & \cr
\hline
C+C & 33 000 & 0.153 &  14 & 0 - 2.0 \cr
\hline
Si+Si & 63 000 &  0.122 & 37 & 0 - 2.6 \cr
\hline
Pb+Pb(6) & 117 000 & 0.57 & 42  & 10.2 - \cr
\hline
Pb+Pb(5) & 59 000 &  0.10 &  88 & 9.1 - 10.2 \cr
\hline
Pb+Pb(4) & 68 000 &  0.10 & 134 & 7.4 - 9.1 \cr
\hline
Pb+Pb(3) & 68 000 &  0.11 &  204 & 5.3 - 7.4 \cr
\hline
Pb+Pb(2) & 45 000 &  0.075 &  281 & 3.4 - 5.3 \cr
\hline
Pb+Pb(1) & 180 000 &  0.05 &  352 & 0 - 3.4 \cr
\hline
\end{tabular}
\end{center}
\caption{Data sets used in analysis. Listed for p+p, C+C, Si+Si and six
centralities of Pb+Pb collisions at 158 $A$GeV are: number of events,
$\sigma/\sigma_{tot}$ - the fraction of the total inelastic cross section 
in that bin,
$\langle N_{W} \rangle$ - the mean number of wounded nucleons, $b$ -
the impact parameter range.}

 \label{data_sets}
\end{table}

\begin{table}[h]
\begin{center}
\begin{tabular}{|c|c|c|c|c|}
\hline
$y_{\pi}$  & $A [\frac{c}{GeV}]$ & $B
[\frac{GeV}{c}]$ &
$C [\frac{deg.^{2}GeV}{c}]$  \cr
\hline
\hline
3.9 - 4.1 & 0 & 0.3 & 6500 \cr
\hline
4.1 - 4.3 & 0 & 0.3 & 5500 \cr
\hline
4.3 - 4.5 & 0 & 0.25 & 4500 \cr
\hline
4.5 - 4.7 & 0 & 0.25 & 3500 \cr
\hline
4.7 - 4.9 & 0 & 0.2 & 2500 \cr
\hline
4.9 - 5.1 & 0.5 & 0.2 & 2500 \cr     
\hline
5.1 - 5.3 & 1.0 & 0.1 & 2500 \cr
\hline
5.3 - 5.5 & 1.5 & 0.1 & 2500 \cr
\hline
\end{tabular}
\end{center}
\caption{The parametrization of the NA49 $y - p_{T}$ acceptance at 158 
$A$GeV for positively charged particles (standard configuration of magnetic 
field). For negatively charged particles one has to redefine the azimuthal 
angle (see text) and then use the same parametrization.}
\label{abc}   
\end{table}

\newpage

\begin{table}[h]
\begin{center}  
\begin{tabular}{|c|c|c|c|c|c|c|}
\hline
  & $ \langle N \rangle $ & $\sigma _N$ & $\overline{p_{T}}$
[MeV/c]
&$\sigma _{p_T}$ [MeV/c]
& $\Phi_{p_{T}}$ [MeV/c] \cr
\hline
\hline
p+p (all) & 1.4 & 1.3 & 304 & 196 & 2.2 $\pm$ 0.3 \cr
\hline
p+p (-) &  0.6  & 0.7 & 283 & 179 & 0.8 $\pm$ 0.1 \cr
\hline
p+p (+) & 0.8  & 0.9 & 317 & 206 & -1.4 $\pm$ 0.3 \cr
\hline
\hline

C+C (all) & 10 & 4.3 & 300 & 210 & 5.4 $\pm$ 0.7 \cr
\hline
C+C (-) &  4.5 & 2.4 & 279 & 190 & 1.8 $\pm$ 0.8 \cr
\hline
C+C (+) &  5.5 & 2.7 & 317 & 224 & 0.7 $\pm$ 0.7 \cr
\hline
\hline

Si+Si (all) & 27 & 7 & 301 & 217 & 4.9 $\pm$ 0.8 \cr
\hline
Si+Si (-) & 12 & 4 & 277 & 195 & 2.6 $\pm$ 0.5 \cr
\hline
Si+Si (+) & 15 & 4 & 320 & 231 & -0.2 $\pm$ 0.7 \cr
\hline
\hline

Pb+Pb(6) (all)  & 39 & 18 & 299 & 220 & 7.2 $\pm$ 0.7 \cr
\hline
Pb+Pb(6) (-) & 18  & 9 & 270 & 195 & 4.5 $\pm$ 0.5 \cr
\hline
Pb+Pb(6) (+) & 21  & 10 & 325 & 237 & 1.9 $\pm$ 0.7 \cr
\hline
\hline

Pb+Pb(5) (all) & 73 & 17 & 305 & 226 & 6.6 $\pm$ 0.7 \cr
\hline
Pb+Pb(5) (-) & 34 & 9 & 273 & 199 &  4.5 $\pm$ 0.7 \cr
\hline
Pb+Pb(5) (+) & 39 & 9 & 333 & 245 & 0.6 $\pm$ 0.8 \cr
\hline
\hline

Pb+Pb(4) (all) & 104 & 19 & 309 & 230 & 5.6 $\pm$ 0.8 \cr
\hline
Pb+Pb(4) (-) & 49 & 10 & 276 & 202 &  3.8 $\pm$ 0.5 \cr
\hline
Pb+Pb(4) (+) & 55 & 11 & 337 & 249 & -0.6 $\pm$ 0.9 \cr
\hline
\hline

Pb+Pb(3) (all) & 148 & 21 & 312 & 233 & 4.6 $\pm$ 0.8 \cr
\hline
Pb+Pb(3) (-) & 69 & 11 & 279 & 204 &  2.9 $\pm$ 0.8 \cr
\hline
Pb+Pb(3) (+) & 79 & 12 & 342 & 252 &  -1.3 $\pm$ 0.8 \cr
\hline
\hline

Pb+Pb(2) (all) & 193 & 21 & 315 & 234 & 2.2 $\pm$ 1.0 \cr
\hline
Pb+Pb(2) (-) & 90 & 11 & 281 & 205 & 2.4 $\pm$ 0.8 \cr
\hline
Pb+Pb(2) (+) & 103 & 13 & 344 & 254 & -3.7 $\pm$ 1.1 \cr
\hline
\hline

Pb+Pb(1) (all) & 230 & 19 & 317 & 236 & 1.4 $\pm$ 0.8 \cr
\hline
Pb+Pb(1) (-) & 108 & 11 & 281 & 203 & 0.9 $\pm$ 0.6 \cr
\hline
Pb+Pb(1) (+) & 122 & 12 & 349 & 257 & -2.9 $\pm$ 0.8 \cr
\hline
\hline

\end{tabular}
\end{center} 
\caption{Measured inclusive and event-by-event parameters for 
accepted particles. $\langle N \rangle $, $\sigma _N$, $\overline{p_{T}}$
and $\sigma _{p_T}$ values are not corrected for acceptance.
$\Phi_{p_{T}}$ values are corrected for limited two track resolution. The
systematic error of $\Phi_{p_{T}}$ is smaller than 1.6 MeV/c.}
\label{data}
\end{table}


\newpage

\newpage

\newpage

\newpage

\newpage

\newpage

\newpage

\newpage

\newpage

\newpage

\newpage

\newpage

\newpage

\newpage

\end{document}